\documentclass[11pt,a4paper]{article}
\pdfoutput=1
\usepackage{jheppub}

\usepackage{graphicx}
\usepackage{mathtools}
\usepackage{color}

\newcommand{\Di}[1]{{\cal D}_{#1}}
\def\beq{\begin{equation}}   
\def\eeq{\end{equation}}
\def\bea{\begin{eqnarray}}  
\def\eea{\end{eqnarray}}

\allowdisplaybreaks[1]

\title{The fully differential hadronic production of a Higgs boson through bottom-quark fusion at NNLO}

\author[a]{Stephan Buehler,}
\author[a]{Franz Herzog,}
\author[a]{Achilleas Lazopoulos}
\author[a]{ and Romain Mueller}

\affiliation[a]{Institute for Theoretical Physics,\\
ETH Zurich,8093 Zurich, Switzerland}

 \emailAdd{buehler@itp.phys.ethz.ch}

 \emailAdd{fherzog@itp.phys.ethz.ch}

 \emailAdd{lazopoli@itp.phys.ethz.ch}

 \emailAdd{muellrom@itp.phys.ethz.ch}

\abstract{
The fully differential computation of the hadronic production cross section of a Higgs boson via bottom quarks is presented at NNLO in QCD. Several differential distributions with their corresponding scale uncertainties are presented for the $8$ TeV LHC.  This is the first application of the method of non-linear mappings for NNLO differential calculations at hadron colliders.
 }


\preprint{}

\begin{document}
\maketitle


\section{Introduction}
\label{sec:introduction}

The Large Hadron Collider is now at its third year of successful operation and both ATLAS and CMS report tantalizing hints of a Higgs boson at about 125 GeV. By the end of the 2012 run the experiments are likely to be able to either confirm those hints as a firm discovery or else  exclude any Standard Model (SM) Higgs boson. In the event of a firm discovery further detailed examination of various production and decay channels will be necessary to determine the nature of the Higgs sector. 

The dominant production channel in the SM, but also in all non-fermiophobic models of new physics, is single Higgs hadroproduction. Within the SM the production mechanism is dominated by gluon fusion, since the alternative mechanism of quark annihilation is severely suppressed by the small Yukawa coupling of bottom and light quarks to the Higgs boson.  However, if the Higgs sector is non-minimal, as is the case in any two-Higgs-doublet model (among which the MSSM is the most studied example) the Yukawa coupling to down-type quarks is enhanced by a factor of $\tan{\beta}$ (the ratio of the vacuum expectation values of the two doublets) and the contribution of the $b\bar{b}\to H$ process becomes significant. Furthermore the production cross section through gluon fusion decreases due to the enhanced, negative top-bottom interference diagrams. In such a scenario, the production of a Higgs boson via $b\bar{b}$  pairs contributes much more than in the SM, and a detailed description of this process is desirable. In other BSM models, for example in models with dynamically generated Yukawa couplings~\cite{Babu:1999me,Giudice:2008uua}, both the bottom and the charm quarks have enhanced couplings to the Higgs boson and charm annihilation becomes important as well.  

The experimental searches are currently focused on measuring an enhanced production rate via bottom annihilation in the  $\tau^+\tau^-$ decay channel with the MSSM as the default BSM model~\cite{Chatrchyan:2012vp,ATLAS_tau_tau}. There are, moreover, several studies on measuring single Higgs decaying to bottom quarks in more generic models in which bottom annihilation is the dominant production channel, using, for example, three $b$-tagged jets~\cite{Kao:2009jv,Baer:2011af}, or measuring the ratio of three heavy ($c$- or $b$-) jet events to three $b$-jet events to discriminate between classes of models with two Higgs doublets~\cite{Atwood:2003yg}.

Bottom quark annihilation has been the subject of much theoretical discussion in the last decade due to the freedom in treating the initial state bottom quarks. Bottom quarks lie in an intermediate mass range between the non-perturbative regime of the proton mass and the typical scale of a hard scattering event at the LHC. One can retain their small mass in the calculation, and exclude them from the proton constituents (four flavor scheme -- 4FS) or treat them as massless partons with their own parton distribution functions (five flavor scheme -- 5FS). In the 4FS the inclusive cross section develops large logarithms $\sim\log({m_b\over m_H})$ due to the collinear production of $b$-quarks which is regulated by the bottom mass. In the 5FS these logarithms are re-summed to all orders by the DGLAP evolution inside the bottom PDFs, for all scales up to the factorization scale adopted in the calculation. Improved convergence of the perturbative expansion is an advantage of the 5FS approach, but at the same time it makes the 5FS prediction very sensitive to the choice of factorization scale. It has been realized that if the factorization scale is set to low values $\sim m_H / 4$, both the 5FS and the 4FS predictions for the inclusive cross sections agree with each other within their respective uncertainties~\cite{Maltoni:2003pn,Boos:2003yi,Plehn:2002vy}, and there is an open discussion as to how one would combine information from both approaches~\cite{Harlander:2011aa,Maltoni:2012pa}.

In the 4FS, the lowest order process would be $gg\to b\bar{b} H$ which begins at order $\alpha_s^2$ in the QCD perturbative expansion and is known at next-to-leading-order (NLO) in QCD~\cite{Dittmaier:2003ej,Dawson:2005vi,Dawson:2004wq,Dawson:2003kb}. The process $bg\to bH$, which starts at order $\alpha_s$, has also been studied at NLO in QCD~\cite{Dawson:2004sh} and with electroweak (EW) corrections~\cite{Dawson:2010yz}. In the 5FS the lowest process is  $b\bar{b}\to H$. Hence the LO 4FS process where a non-collinear bottom pair is observable, is only reached for the first time at NNLO in the 5FS. The inclusive cross section, in the 5FS, of  $b\bar{b}\to H$ is known at NNLO in QCD~\cite{Harlander:2003ai} as well as at NLO in EW~\cite{Dittmaier:2006cz}. NNNLO threshold re-summed soft and collinear terms are also known~\cite{Kidonakis:2007ww} and the transverse momentum distribution of the Higgs boson has been studied with re-summation techniques~\cite{Field:2004nc,Belyaev:2005bs} . Also known at NNLO are the zero-, one- and two-jet rates and related distributions~\cite{Harlander:2011fx}, quantities which can be obtained already from the differential $H+\text{jet}$ computation at NLO~\cite{Harlander:2010cz} in combination with the fully inclusive NNLO cross section.

In this paper we present the fully differential NNLO QCD cross section for $b\bar{b}\to H$ in the 5FS within the SM. NLO computations are currently performed with very well automated methods. Obtaining fully differential cross sections and decay rates at one order  higher in the perturbative  expansion 
requires the solution of new challenging problems. Regarding the treatment of the real emissions, pioneered for NLO computations in~\cite{Ellis:1980wv, Mele:1990bq}, rapid progress has been made in the last decade~\cite{Kosower:1997zr,GehrmannDeRidder:2003bm,GehrmannDeRidder:2005cm, Daleo:2006xa, GehrmannDeRidder:2009fz, Daleo:2009yj, Glover:2010im, Boughezal:2010mc, Abelof:2011jv, Gehrmann:2011wi, Weinzierl:2003fx, Frixione:2004is, Somogyi:2005xz, Somogyi:2006da, Somogyi:2006db, Bolzoni:2009ye, Bolzoni:2010bt, Somogyi:2008fc,Aglietti:2008fe,Czakon:2010td,Czakon:2011ve,Boughezal:2011jf,Anastasiou:2003gr,Catani:2007vq, Binoth:2000ps,Binoth:2004jv}, mainly focusing on the treatment of the double real emission\footnote{A variety of methods has been proposed covering the range from fully orthodox to outright heretic.}, which resulted in the fully differential calculations of Higgs production via gluon fusion~\cite{Anastasiou:2002qz,Anastasiou:2004xq,Anastasiou:2005qj,Grazzini:2008tf}, Drell-Yan~\cite{Anastasiou:2003ds,Anastasiou:2003yy,Melnikov:2006kv,Catani:2009sm,Gavin:2012kw}, associated Higgs production with a vector boson~\cite{Ferrera:2011bk}, three jet production from $e^+e^-$~\cite{Anastasiou:2004qd,GehrmannDeRidder:2004tv,GehrmannDeRidder:2007hr,GehrmannDeRidder:2007jk} and diphoton production~\cite{Catani:2011qz}.   

However, further development of methods and new ideas are necessary for efficient cancellations of infrared singularities and  
evaluations of novel two-loop amplitudes in more complicated LHC processes. 

With this paper we also take the opportunity to complete the second NNLO application, after the fully differential decay $H \to b\bar b$ \cite{Anastasiou:2011qx}, 
using the method of nonlinear mappings to factorize singularities in the double real corrections \cite{Anastasiou:2010pw}. The double real contributions have often been regarded as the bottleneck of NNLO, and this paper therefore also demonstrates the validity of the approach as a method for NNLO corrections in hadronic collisions. 

The paper is organized as follows: in section~\ref{sec:setup} we set up the notation and describe the main components of the calculation. In section~\ref{soft-hard} we provide some detail about the treatment of the separation of  soft and hard contributions. In section~\ref{sec:calculational_details} we describe the treatment of the double real and the real-virtual components. In section~\ref{collinear} we present the way we perform the (non-trivial at NNLO) convolutions for the collinear subtraction terms in mass factorization. In section~\ref{sec:numerical_results} we provide various numerical results both on jet rates, $p_T$ and rapidity distributions; demonstrate the completely differential nature of our calculation and provide typical results for the case in which the Higgs boson decays to two photons, including standard experimental cuts on photon momenta and isolation.

\section{Notational setup and conventions}
\label{sec:setup}
\subsection{Fully differential calculations}
One of the merits of fully differential calculations is  the possibility to arrive at theoretical predictions for observables in the presence of final state phase-space cuts, like those used in experimental analyses, under the precondition that the observable defined is infra-red safe. 
Throughout this article the dependence on such arbitrary phase-space constraints will be contained in the 
jet-function $\mathcal J (\{p\}_f)$, where $\{p\}_f$  denotes the set of final state momenta in the lab frame.
We will refer to the fully differential cross section as $\sigma[\mathcal J]$,
which we schematically define as
\beq
\sigma[\mathcal J] = \sum_f \int d\sigma_f \, \mathcal J (\{p\}_f) \, ,
\eeq
where the  sum is over all final states $f$. 

The usual role of the jet function $\mathcal{J}$ is to apply arbitrary final state phase-space cuts while ensuring infra-red safety. Here we promote it to a further task, which is to keep track of the bin-integrated cross section for  any given differential observable with or without applying phase-space cuts. This can be achieved simply at the level of Monte Carlo integration by passing to $\mathcal J$ not only the set of final state momenta but also the weight of the given event. The role of the jet function becomes crucial in all amplitudes that have soft and collinear singularities which are regulated by counter terms. In such cases the jet function is keeping track of the kinematics of every subtraction term.

\subsection{Hadronic cross section}
We consider the following hadronic process  
\beq
P_1+P_2 \to H+X \, ,
\eeq
where $P_1$, $P_2$ are the incoming hadrons, $H$ denotes the Higgs boson and $X$ generically denotes surplus QCD radiation in the final state. 
The Higgs boson is assumed to couple only to bottom quarks via the SM Yukawa interaction. 
Assuming the usual factorization, the fully exclusive hadronic cross section can be written as  
\begin{eqnarray}
 \sigma_{P_1P_2\to H+X}\left[ \mathcal{J}\right]  & = &\sum_{i_1,i_2} \int_0^1 dx_1 dx_2 \, \theta(x_1x_2-\tau)f_{i_1}(x_1)f_{i_2}(x_2)\sigma_{i_1i_2\to HX}[\mathcal{J}] \, ,
\label{master}
\end{eqnarray}
where the $f_{i}(x)$ denote the bare (unrenormalized) parton distribution functions (PDFs) in the 5FS, 
$x_1$ and $x_2$ are the usual Bjorken-$x$ momentum fractions of the partons $i_1$ and $i_2$ respectively, and $\tau=\frac{m_H^2}{S}$, where $m_H^2$ is the (on-shell) mass of the Higgs-boson and $S$ is the square of the total center of mass  (CoM) energy of the colliding hadrons.
By $\sigma_{i_1i_2\to HX}$ we denote the partonic cross section for the processes
\beq
i_1(p_1)+i_2(p_2) \to H(p_H)+X(i_3(p_3),i_4(p_4),\ldots)\, , \qquad i_{1,2,3,\ldots} \in  \{\bar b,\bar q ,g,q,b\}.
\eeq
The PDFs we have inserted in eq.(\ref{master}) are bare and  we still have to rewrite them in terms of the renormalized 
PDFs. This step will introduce collinear counter terms that  cancel the  initial state collinear singularities of the partonic cross section, which remain after all real and virtual corrections are added together. This cancellation is achieved fully numerically in our calculation. 
We  outline the way collinear counter terms  can be computed process-independently in section \ref{collinear}. \\

\begin{figure}[t]
	\center
	\includegraphics[scale=0.8]{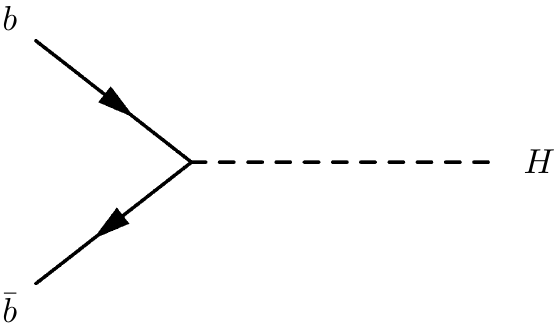}
	\caption{\textit{\small{ The leading order  contribution to $b \bar b \to H$, see section \ref{partonic}. }}}
	\label{fig:born}
\end{figure}

\subsection{Partonic cross sections}
\label{partonic}
Expanding the partonic cross section to NNLO in QCD we obtain
\begin{eqnarray}
\sigma_{ij\to HX} [\mathcal J] &=&y_b^2\Bigg[ \sigma_{ij\to H}^B[\mathcal J]+\frac{\alpha_s}{\pi}\left(\sigma_{ij\to Hk}^R[\mathcal J]+\sigma_{ij\to H}^V[\mathcal J]\right) \nonumber \\
& &\qquad +\left(\frac{\alpha_s}{\pi}\right)^2\left(\sigma_{ij\to Hkl}^{RR}[\mathcal J]+\sigma_{ij\to Hk}^{RV}[\mathcal J]+\sigma_{ij\to H}^{VV}[\mathcal J]\right)  +\mathcal{O}\left(\alpha_s^3 \right )     \Bigg],
\end{eqnarray}
where a sum is implied over all final state flavors $k$ and $l$ leading to a possible subprocess. Here $y_b=y_b(\mu)$ and $\alpha_s=\alpha_s(\mu)$ are the $\overline{\mathrm{MS}}$ renormalized bottom Yukawa  and strong couplings, with 5 active flavors. 
We set the dimensional regularization scale $\mu$ to be equal to both the renormalization and factorization scales, $\mu_R$ and $\mu_F$. 
Separation of the two scales can be easily achieved via the relations given in appendix \ref{scaleseparation}.

\begin{figure}[h!]
	\center
	\includegraphics[scale=0.8]{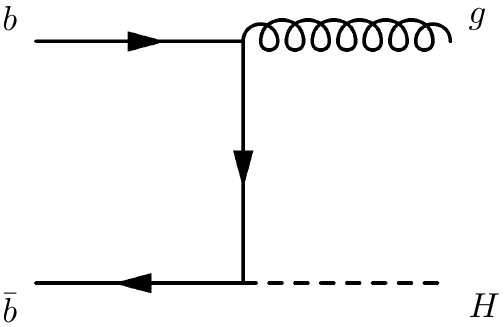}
	\hspace{1cm}
	\includegraphics[scale=0.8]{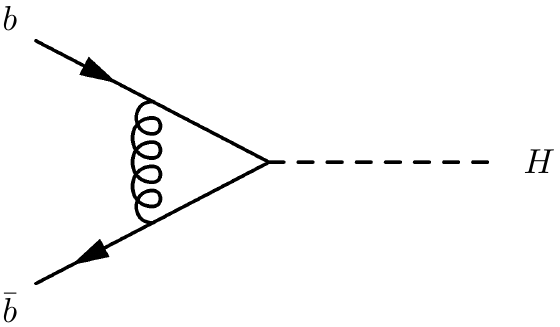}
	\caption{\textit{\small{ Some of the diagrams contributing to $b \bar b \to H$ at NLO. These contributions are denoted as \emph{real} (left) and \emph{virtual} (right). }}}
	\label{fig:NLO}
\end{figure}

\begin{figure}[h!]
	\center
	\includegraphics[scale=0.8]{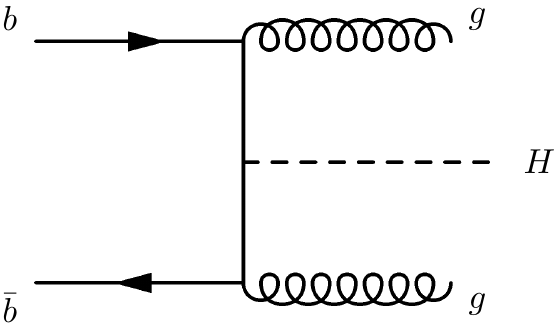} \hfill
	\includegraphics[scale=0.8]{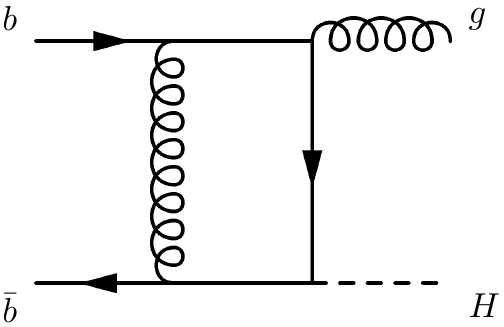} \hfill
	\includegraphics[scale=0.8]{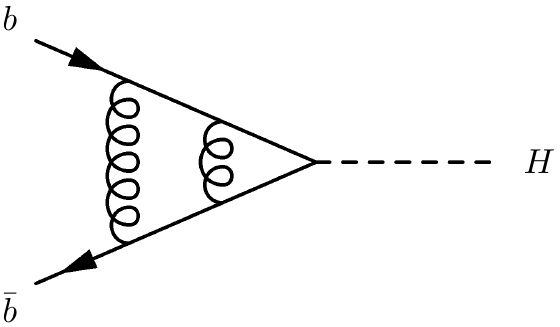} \hfill
	\caption{\textit{\small{ Some of the diagrams contributing to $b \bar b \to H$ at NNLO. These   contributions are denoted as \emph{double real} (left), \emph{real-virtual} (center) and \emph{double virtual} (right).}}}
	\label{fig:NNLO}
\end{figure}

The leading order  contribution, fig.~\ref{fig:born}, is denoted by $\sigma_{ij\to H}^B$. At  NLO there are two separate contributions (see fig.~\ref{fig:NLO}):
\begin{itemize}
\item The \textit{real} ($\sigma_{ij\to Hk}^{R}$):\\
Corresponding to the emission of an extra  particle $k$. 
\item The \textit{virtual} ($\sigma_{ij\to H}^{V}$):\\
Corresponding  to the emission and re-absorption of a  virtual  particle. 
\end{itemize}
At NNLO there are three separate contributions (see fig.~\ref{fig:NNLO}):
\begin{itemize}
\item The \textit{double real} ($\sigma_{ij\to Hkl}^{RR}$):\\
Corresponding  to the emission of two  particles $k$ and $l$. 
\item The \textit{real-virtual} ($\sigma_{ij\to Hk}^{RV}$):\\
Corresponding  to the emission of one  particle $k$ as well as the emission and re-absorption of a virtual  particle. 
\item The \textit{double virtual} ($\sigma_{ij\to H}^{VV}$):\\
Corresponding to the emission and re-absorption of two virtual  particles. 
\end{itemize}
Real and virtual corrections suffer from \textit{infra-red} as well as \textit{ultra-violet} divergences. 
We use conventional dimensional regularization with $d=4-2\epsilon$ to regularize such divergences, which then appear as poles in $\epsilon$.
More specifically \textit{ultra-violet} divergences present in virtual corrections are absorbed into 
the renormalized couplings, see e.g. \cite{Anastasiou:2011qx,Harlander:2003ai} or any textbook on QCD.
In contrast, \textit{infra-red} divergences cancel only after summing all real and virtual corrections contributing to a given infra-red safe observable.

Factorized singularities on the unit hypercube may be dealt with using the plus-distribution expansion
\beq
x^{-1-n\epsilon} = \frac{\delta(x)}{-n\epsilon} + \sum_{m=0}^\infty \frac{(-n\epsilon)^m}{m!} \Di{m}(x) \, ,
\label{plus}
\eeq
where $\Di{m}(x)=\left[\frac{\ln^m(x)}{x}\right]_+$ and the plus-distribution is defined through
\beq
\int_0^1 dx \, f(x) \left[\frac{g(x)}{x}\right]_+=\int_0^1 dx \, g(x) \left(\frac{f(x)-f(0)}{x}\right).
\eeq 
Beyond NLO a factorization of singularities is highly non-trivial. In this work it is
 achieved systematically using the method of nonlinear mappings \cite{Anastasiou:2010pw}.

Care must be taken when dealing with infrared singularities of real emission amplitudes:  the plus-distribution, eq.(\ref{plus}),  also acts 
on the jet function, such that cancellations happen also at the differential level and are therefore fully local. \\

We now give a brief overview of the matrix elements and phase-space measures required for the computation of the partonic cross section.
Here we assume the amplitudes to be color and spin summed. Averaging and phase-space symmetry factors will be explicitly factored out. 
\begin{itemize}
\item[i)] Purely virtual corrections:\\
The purely virtual  corrections include the Born, virtual and double virtual contributions and are of the form
\beq
\sigma_{b\bar b\to H}[\mathcal{J}]=\frac{N_{b\bar b}}{2s_{12}} \int d\Phi_{2\to 1} |M_{b\bar b \to H}|^2 \mathcal{J}(p_1,p_2,p_H) \, ,
\eeq
where $s_{12} = (p_1+p_2)^2$ is the partonic CoM energy and $N_{b\bar b}=1/36$. The corresponding  phase-space volume element is trivially given by
$d\Phi_{2\to 1} = 2\pi\delta(s_{12}-m_H^2)$, constraining $s_{12} = m_H^2=p_H^2$. Regarding the computation of amplitudes 
we refer the reader to \cite{Anastasiou:2011qx} where the full virtual matrix-elements can be found.
Explicit expressions for these contributions are given in section \ref{S}.

\item[ii)] Single real emissions:\\
The single real emissions include real and real-virtual corrections and are of the form
\beq
\sigma_{i_1 i_2\to H i_3}[\mathcal{J}] = \frac{N_{i_1i_2}}{2 s_{12}} \int d\Phi_{2\to 2} |M_{i_1i_2 \to Hi_3}|^2 \mathcal{J}(p_1,p_2,p_3,p_H)\, .
\eeq
Further details are given in section \ref{R}.
 
\item[iii)]Double real emissions:\\
These are of the form
\beq
\sigma_{i_1 i_2\to H i_3i_4}[\mathcal{J}] = \frac{N_{i_1i_2}}{2s_{12}} \int d\Phi_{2\to 3} |M_{i_1i_2 \to Hi_3i_4}|^2 \mathcal{J}(p_1,p_2,p_3,p_4,p_H) \, .
\eeq
Further details are given in section \ref{RR}.
\end{itemize}


\section{Separation of soft and hard}
\label{soft-hard}
Since in the soft limit of a $2\to 1$ process the produced particle is at rest in the partonic center of mass frame, there is no 
difference between the soft piece of a fully differential partonic cross section and that of a fully inclusive partonic cross section.
It is therefore very convenient to isolate the soft contribution ($\sigma_S$) to the partonic cross section ($\sigma$) from the hard one ($\sigma_H$), i.e.
\beq
\sigma=\sigma_S+\sigma_H.
\eeq 
This allows for a fully analytic treatment of $\sigma_S$, while $\sigma_H$ must, as far as external kinematics are concerned, be treated numerically.    
Let us introduce the variable
\beq
z=\frac{m_H^2}{s_{12}} \, .
\eeq 
Then the soft limit of all real emission amplitudes corresponds to $z\to1$, which identifies  the production threshold.
Given that infrared singularities are of logarithmic nature, the divergence at $z=1$ can be exposed as follows
\beq
\sigma(z)[\mathcal{J}] = \delta(1-z) \widetilde \sigma_V (\epsilon) [\mathcal{J}]|_{z=1} + \sum_n \frac{\widetilde{\sigma}^{(n)}_R (z, \epsilon)[\mathcal{J}]}{(1-z)^{1+n\epsilon}} \, ,
\label{sigmaz}
\eeq
where $\widetilde{\sigma}_{V}$ denotes the purely virtual correction, while $\widetilde{\sigma}^{(n)}_R(z,\epsilon)$  denotes real corrections
collectively (at NNLO this  includes both real-virtual as well as double real corrections).

Separation into soft and hard parts can now be achieved by adding and subtracting the soft limit from the second term in the above, 
yielding
\begin{align}
\sigma(z)[\mathcal{J}]&=\underbrace{\delta(1-z)\widetilde{\sigma}_{V}(\epsilon)[\mathcal{J}]|_{z=1}+\sum_{n}\frac{\widetilde{\sigma}^{(n)}_R(1,\epsilon)[\mathcal{J}]|_{z=1}} {(1-z)^{1+n\epsilon}}}_{\equiv\sigma_S} \nonumber \\
& \quad +\quad \, \underbrace{\sum_{n}\frac{\widetilde{\sigma}^{(n)}_R(z,\epsilon)[\mathcal{J}]-\widetilde{\sigma}^{(n)}_R(1,\epsilon)[\mathcal{J}]|_{z=1}}{(1-z)^{1+n\epsilon}}}_{\equiv\sigma_H} \, ,
\label{sigmazHS}
\end{align}
such that $\sigma_H$ is integrable in the range $z \in [\tau,1]$. Of course this decomposition of the partonic cross section into its soft and hard components 
is not  unique: one could use any other subtraction term with the correct limit, thereby including, for example, the luminosity function. Our choice, however, has the nice property that the soft part $\sigma_S$ can be expanded purely in terms of $\delta$- and 
plus-distributions via eq.(\ref{plus}), 
$$
\sigma_S[\mathcal{J}]=c_0\delta(1-z)\, \mathcal{J} |_{z=1}+\sum_{n=0}^\infty c_n \, \Di{n}(1-z) \, \mathcal{J}|_{z=1} \, .
$$
Thereby all threshold divergences between $\widetilde{\sigma}_{V}$ and $\widetilde{\sigma}^{(n)}_R$
are canceled analytically, leaving only a finite threshold contribution. Furthermore this framework provides a natural way to incorporate 
threshold re-summation in  fully differential calculations.

\section{Details on the calculation}
\label{sec:calculational_details}
\subsection{The single real} 
\label{R}
The single real partonic cross section may be expressed as 
\begin{equation}
\sigma^{R}_{ij}\left[ \mathcal{J}\right] = y_b^2 \left(\frac{\alpha_s}{\pi}\right)  \frac{N_{ij}}{2s_{12}} \int d\widetilde\Phi_2  |M^{R}_{ij}|^2 \mathcal{J}(p_1,p_2,p_3,p_H) \, .
\end{equation}
We define $d\widetilde\Phi_2$ to be the conventional phase-space volume $d\Phi_2$ up to some renormalisation constants.
Here we have to consider 6 separate channels:
\[
N_{ij}|M^{R}_{ij}|^2 = 
\left\{ 
\renewcommand{\arraystretch}{1.7}
\begin{array}{ll} 
 \frac{1}{2^2\cdot3^2} |M^{R}_{b\bar b \to gH}|^2       &\quad\mbox{if  $(i,j) \in \{(b,\bar b),(\bar b,b) \}$}\,; \\
 \frac{1}{2\cdot(2-2\epsilon)\cdot3\cdot8} |M^{R}_{b g \to bH}|^2 &\quad \mbox{if  $(i,j) \in  \{(b,g),(\bar b,g)\}$}\,;\\ 
 \frac{1}{2\cdot(2-2\epsilon)\cdot3\cdot8} |M^{R}_{g b \to bH}|^2  &\quad \mbox{if  $(i,j) \in  \{(g,b),(g,\bar b)\}$}\,.\\  
\end{array} 
\right. 
\]
The corresponding amplitudes may all be found in \cite{Anastasiou:2011qx}.
A convenient phase space parametrization is given by
\beq
\label{phi2}
d\Phi_{2}=\frac{1}{8\pi} \frac{(4\pi)^\epsilon}{\Gamma(1-\epsilon)} s^{-\epsilon}(1-z)^{1-2\epsilon}  \, [\lambda (1-\lambda)]^{-\epsilon}\,d\lambda,
\eeq
where $\lambda \in [0,1]$, with the Lorentz invariants taking the simple form
\begin{eqnarray}
&& s_{13}=(p_1-p_3)^2=-s_{12}(1-z)\lambda \, , \nonumber \\ 
&&  s_{23}=(p_2-p_3)^2=-s_{12}(1-z)(1-\lambda) \, .
\label{eq:realparm}
\end{eqnarray}
Note that  the singularities of $s_{13}$ and $s_{23}$ are factorized in $\lambda$, $(1-\lambda)$ and $(1-z)$ which
allows for a simple subtraction of the poles using eq.(\ref{plus}). This also allows us to identify 
\beq 
\sigma^{R}_{b\bar b}\left[ \mathcal{J}\right]=\frac{ \widetilde{\sigma}^R_{b\bar b}\left[ \mathcal{J}\right]}{(1-z)^{1+2\epsilon}} \, .
\eeq
The calculation of the hard part then trivially follows from eq.(\ref{sigmazHS}).

\subsection{The real-virtual}
\label{RV}
The real-virtual partonic cross section may be expressed as 
\[
\sigma^{RV}_{ij}\left[ \mathcal{J}\right] = (y_b)^2 \left(\frac{\alpha_s}{\pi}\right)^2  \frac{N_{ij} }{2s_{12}}  \int d\widetilde{\Phi}_2\, 2\mathrm{Re}\left\{ M^{RV}_{ij} {M^{R}_{ij}}^* \right\} \mathcal{J}(p_1,p_2,p_3,p_H)\,,
\]
where we have taken the liberty to define $d\widetilde{\Phi}_2$ to equal eq.(\ref{phi2}) up to some renormalization constants. 
Then 
\[
N_{ij}2\mathrm{Re} \left\{M^{RV}_{ij}{M^{R}_{ij}}^*\right\}= 
\left\{ 
\renewcommand{\arraystretch}{1.7}
\begin{array}{ll} 
 \frac{1}{2^2\cdot3^2} \,2\mathrm{Re}\left\{M^{RV}_{b\bar b \to gH}{\left (M^{R}_{b\bar b \to gH}\right)}^*\right\}             &\;\: \mbox{if  $(i,j) \in \{(b,\bar b),(\bar b,b) \}$}\,; \\
 \frac{1}{2\cdot(2-2\epsilon)\cdot3\cdot8} \,2\mathrm{Re}\left\{M^{RV}_{bg \to bH}{\left (M^{R}_{b g \to bH}\right)}^*\right\}  &\;\: \mbox{if  $(i,j) \in  \{(b,g),(\bar b,g)\}$}\,;\\ 
 \frac{1}{2\cdot(2-2\epsilon)\cdot3\cdot8} \,2\mathrm{Re}\left\{M^{RV}_{gb \to bH}{\left (M^{R}_{g b \to bH}\right)}^*\right\}  &\;\: \mbox{if  $(i,j) \in  \{(g,b),(g,\bar b)\}$}\,.\\  
\end{array} 
\right. 
\]
The real-virtual amplitude can be obtained from the corresponding one from the decay process $H\to b\bar{b}$ published in \cite{Anastasiou:2011qx}  by crossing particles to the initial state.
The box integrals we encounter in this amplitude are entirely expressible in terms of Gauss' hypergeometric function ${}_2F_1(1,-\epsilon,1-\epsilon,z)$  
where $z$ can be in any of the three sets $S_{fine}$, $S_{inv}$ and $S_{nl}\,$:
\begin{align}
S_{fine} &=\left\{\frac{-s_{13}}{s_{12}}, \frac{-s_{23}}{s_{12}}\right\},\qquad \;\;\, S_{inv} =\left \{\frac{-s_{12}}{s_{13}}, \frac{-s_{12}}{s_{23}} \right\}, \nonumber \\
S_{nl} &= \left\{\frac{-s_{13}}{s_{23}}, \frac{-s_{23}}{s_{13}}, \frac{-s_{12} m_H^2}{s_{13}s_{23}}, \frac{-s_{13}m_H^2}{s_{23}s_{12}}, \frac{-s_{23}m_H^2}{s_{12}s_{13}}\right\}. 
\end{align}

When attempting a direct subtraction of the singularities created by the real emission, the points of subtraction overlap with singular points of the hypergeometric functions in the box integrals.  
It was found in \cite{Anastasiou:2011qx} that one can apply transformations on the argument of the functions to circumvent this difficulty.
Since here we are no longer in the euclidean regime of this amplitude, the required  transformations
are different than in~\cite{Anastasiou:2011qx}. Analyzing integral representations, we find that we have to apply the following identities:
\begin{itemize}
\item If $z \in S_{fine}$ the soft-collinear limits are well defined.
\item If $z \in S_{nl}$ we apply
$$
{}_2F_1(a,b;c;z)\mapsto(1-z)^{-b}{}_2F_1\left(c-a, b; c; \frac{z}{z-1}\right). 
$$
\item If $z \in S_{inv}$ we employ the argument inversion, 
\begin{align}
{}_2F_1(a,b;c;z) \mapsto & \quad\: \frac{\Gamma  \left( b-a \right) \Gamma 
 \left( c \right) {}_2F_1 \left( a,a-c+1;a-b+1;\frac{1}{z} \right) }
 {\Gamma  \left( b \right) \Gamma  \left( c-a
 \right) (-z)^{a}} \nonumber \\
&+\frac {\Gamma  \left( a-b \right) \Gamma  \left( c
 \right) {}_2F_1 \left( b,b-c+1;b-a+1;\frac{1}{z} \right)  }{\Gamma  \left( a \right) \Gamma  \left( c-b \right)(-z)^{b}} \,.
\end{align}
\end{itemize}
After these transformations are applied, the singularities corresponding to the real emission are factorized in $\lambda$, $(1-\lambda)$ and $(1-z)$.
The soft singularity structure of the real-virtual may then be extracted as 
\beq
\sigma^{RV}_{ij}\left[ \mathcal{J}\right]=\sum_{m=2}^4 \frac{\widetilde{\sigma}^{(m)RV}_{ij}\left[ \mathcal{J}\right]}{(1-z)^{1+m\epsilon}}\,.
\eeq
In the soft limit only the $m=2,4$ coefficients survive and the integration over $\lambda$ can be done analytically. The explicit expressions for the soft limit can be found in   
 appendix \ref{Treshholdexplicit}.
 
The computation of the hard part then follows from eq.(\ref{sigmazHS}). While the structure is more complicated than in the case of the single real, 
a direct subtraction via eq.(\ref{plus}) can still be achieved in a straightforward manner. 
In order to obtain the final Laurent expansion in $\epsilon$ we employ the library \verb HypExp  \cite{Huber:2007dx} to 
expand the hypergeometric functions in terms of polylogarithms.\\

\subsection{The double real}
\label{RR}

The double real partonic cross section can be written as
$$
\sigma^{RR}_{ij}\left[\mathcal{J}\right] = (y_b)^2 \left(\frac{\alpha_s}{\pi}\right)^2  \frac{N_{ij} }{2s_{12}}\int \widetilde{d\Phi}_3 |M^{RR}_{ij}|^2 \mathcal{J}(p_1,p_2,p_3,p_4,p_H) \, ,
$$
where $\widetilde{d\Phi}_3$ is equal to the conventional three-particle phase space element $d\Phi_3$ up to renormalization constants.
Using the discrete symmetries of the squared amplitudes we are able 
to considerably reduce the number of independent channels, which one has to implement separately. 
These symmetries are due to the charge invariance of all the $b\bar b \rightarrow H$ double real amplitudes 
(exchanging $q\leftrightarrow \bar q$ or $b\leftrightarrow \bar b$ leaves the amplitudes invariant).
This leaves us with the following list of channels
\beq
N_{ij}|M^{RR}_{ij}|^2= 
\left\{
\renewcommand{\arraystretch}{1.4}
\begin{array}{ll} 
 \frac{1}{2^2\cdot3^2} \left[ \frac{1}{2!}|M_{b\bar b\rightarrow ggH}|^2 +|M_{b\bar b\rightarrow b\bar bH}|^2\right. & \\
\left.\quad\quad\,\:\:+|M_{b\bar b\rightarrow q\bar qH}|^2 \right] &\mbox{if  $(i,j) \in \{(b,\bar b),(\bar b,b) \}$}\,; \\
 \frac{1}{2^2\cdot3^2} |M_{q\bar q\rightarrow b\bar bH}|^2  & \mbox{if  $(i,j) \in \{ (q,\bar q),(\bar q,q)\}$}\,;\\
 \frac{1}{2^2\cdot3^2} |M_{qb\rightarrow qbH}|^2  & \mbox{if  $(i,j) \in \{(q,b),(\bar q,b),(q,\bar b),(\bar q,\bar b)\}$}\,;\\
 \frac{1}{2^2\cdot3^2} |M_{bq\rightarrow bqH}|^2  & \mbox{if  $(i,j) \in \{(b,q),(\bar b,q),(b,\bar q),(\bar b,\bar q)\}$}\,;\\
 \frac{1}{2\cdot(2-2\epsilon)\cdot3\cdot8} |M_{bg\rightarrow bgH}|^2  & \mbox{if  $(i,j) \in  \{(b,g),(\bar b,g)\}$}\,;\\ 
 \frac{1}{2\cdot(2-2\epsilon)\cdot3\cdot8} |M_{gb\rightarrow bgH}|^2  & \mbox{if  $(i,j) \in  \{(g,b),(g,\bar b)\}$}\,;\\  
 \frac{1}{(2-2\epsilon)^2\cdot8^2} |M_{gg\rightarrow b\bar bH}|^2  & \mbox{if  $(i,j) = (g,g)$}\,;\\   
 \frac{1}{2^2\cdot3^2} \frac{1}{2!}|M_{b b\rightarrow bbH}|^2  & \mbox{if  $(i,j) \in\{(b,b),(\bar b,\bar b)\}$}\,.\\
\end{array}
\right. 
\eeq
By crossing partons from the initial to the final state, we can obtain all of the above from the three amplitudes
$$
|M_{b\bar bggH\rightarrow  0}|^2,\;|M_{b\bar b b\bar bH\rightarrow  0}|^2\; \text{and}\;|M_{b\bar b q\bar qH\rightarrow 0}|^2
$$
published in \cite{Anastasiou:2011qx}. In order to deal with the intricate singularities, their factorization and 
subtraction, we refer the reader to the methods developed in \cite{Anastasiou:2010pw}, which we have implemented faithfully.
As in the single real emissions, the double soft singularity occurs at the threshold. Its structure may be identified as 
\beq
\sigma^{RR}_{b\bar b}\left[ \mathcal{J}\right]=\frac{ \widetilde{\sigma}^{RR}_{b\bar b}\left[ \mathcal{J}\right]}{(1-z)^{1+4\epsilon}} \, .
\eeq


\subsection{The soft} \label{S}
Let us expand $\sigma_S$ in the strong coupling
\beq
\sigma_S [\mathcal J]=\mathcal{B}\cdot \left( \delta (1-z)+\frac{\alpha_s}{\pi} \Delta_{S,\mathit{NLO}}+ \left(\frac{\alpha_s}{\pi}\right)^2 \Delta_{S,\mathit{NNLO}}  +\mathcal{O}(\alpha_s^3) \right) \, \mathcal J |_{z=1} \, ,
\eeq
where 
\beq
\mathcal{B}=\frac{\pi y_b^2}{6m_H^2}\,. 
\label{equ:B}
\eeq
 The NLO correction $\Delta_{S,\mathit{NLO}}$ may be expressed as
\begin{eqnarray}
\Delta_{S,\mathit{NLO}} &=& 
\frac{1}{
\epsilon}\left( -\frac{2}{3}\delta \left( 1-z
 \right) -{\frac {8}{3}}\Di{{0}} \left( 1-z \right)  \right) +  \left( {\frac {8}{3}}\zeta_2-\frac{4}{3} \right) \delta \left( 1-z
 \right) 
 \nonumber\\
&&
 +{\frac {8}{3}}\Di{{0}} \left( 1-z \right) {\it l_H}+{\frac {
16}{3}}\Di{{1}} \left( 1-z \right)
 + \mathcal{O}(\epsilon)\, 
\end{eqnarray}
while the NNLO correction $\Delta_{S,\mathit{NNLO}}$ can be expanded as follows 
\beq
\Delta_{S,\mathit{NNLO}}=\sum_{n}\Delta_{S,\mathit{NNLO}}^{(n)}\epsilon^{n} + \mathcal{O}(\epsilon) \, ,
\eeq
with the only non-vanishing contributions \cite{Anastasiou:private}
\begin{align}
\Delta_{S,\mathit{NNLO}}^{(0)} &=  \bigg(  \left( {\frac {2}{27}} -{\frac {10}{27}}\zeta_2 + \frac{2}{3}\zeta_3 \right) { n_f}-{\frac {64}{9}} \zeta_2 l_H^{2}+ \left( -{\frac {17}{3}}+\frac{8}{3}
\zeta_2+{\frac {250}{9}}\zeta_3  \right) { l_H}\nonumber\\
&\quad
+ {\frac {211}{18}} +{\frac {58}{9}}\zeta_2-{\frac {26}{3}}
\zeta_3 -{\frac {17}{6}}\zeta_4 \bigg) \delta \left( 1-z \right) \nonumber\\
&\quad
 + \bigg(  \left( {\frac {56}{81}}-{\frac {20}{27}}{ l_H}+\frac{2}{9}{{ l_H}}^{2}-{
\frac {8}{9}}\zeta_2 \right) { n_f} - {{ l_H}}^{2}+ \left( {\frac {70}{9}} - {\frac {164}{9}}\zeta_2 \right) { l_H} \nonumber\\
&\quad
- {\frac {212}{27}} + {\frac {20}{3}}\zeta_2 + {
\frac {638}{9}}\zeta_3 \bigg) \Di{{0}} \left( 1-z \right)\nonumber\\
&\quad
+ \left(  \left( {\frac {8}{9}}{ l_H}-{\frac {40}{27}} \right) {
 n_f}+{\frac {68}{3}}+{\frac {128}{9}}{{ l_H}}^{2}-4{ l_H} - 72 \zeta_2 \right) \Di{{1}} \left( 1-z \right)\nonumber\\
&\quad
 + \left( -4+{\frac {8}{9}}{ n_f}+{\frac 
{128}{3}}{ l_H} \right) \Di{{2}} \left( 1-z \right) +{\frac {896}{27}}\Di{{3}} \left( 1-z \right), \\ ~\nonumber \\
\Delta_{S,\mathit{NNLO}}^{(-1)} &=
\left(  \left( \frac{1}{36}+\frac{2}{9}\zeta_2 \right) { n_f}+{\frac {64}{9}}\zeta_2 { l_H}+{\frac {43}{24}}-{\frac {23}{3}}\zeta_2-{\frac {125}{9}}
\zeta_3  \right) \delta \left( 1-z \right)\nonumber\\
&\quad + \left( {\frac {10}{27}}{ n_f}-\frac{16}{3}{ l_H}-{\frac {35}{9}} +{\frac {82}{9}}\zeta_2 \right) 
\Di{{0}} \left( 1-z \right) \nonumber\\
&\quad - \left( {\frac {32}{3}}+{\frac {128}{9}}{ l_H} \right) \Di{{1}} \left( 1-z \right) -{\frac {64}{3}}\Di{{2}} \left( 1-z \right) \\
\intertext{and}
\Delta_{S,\mathit{NNLO}}^{(-2)} &=
 \left({\frac {19}{4}} - \frac{1}{6}{ n_f} - {\frac {32}{9}}\zeta_2 \right) \delta \left( 1-z \right) + \left( 9-\frac{2}{9}{ n_f} \right) \Di{{0}} \left( 1-z \right) \nonumber\\
&\quad
+{\frac {64}{9}}\Di{{1}} \left( 1-z \right) \, .
\end{align}
Here $n_f$ is the number of light flavors, $\zeta_n$ are the usual Riemann zeta values and 
\beq \label{lH}
l_H=\log\left(\frac{m_H^2}{\mu^2}\right) \, .
\eeq
The explicit soft limits of the real-virtual and double real pieces that are included in $\sigma_S$ can be found separately, and with their full color-dependence, in appendix \ref{Treshholdexplicit}.

\section{Collinear factorization} 
\label{collinear}
\label{collinearintegrals}

Parton distribution functions are renormalized to absorb initial state collinear
singularities via
\begin{equation}
\tilde{f}_i(z, \mu) = \left( \Gamma_{ij}( \mu) \otimes f_j \right)(z) \, , 
\label{eq:renormed_pdf}
\end{equation}
where $\mu$ is the factorization scale and $f_i$ are the
bare parton densities. In the following discussion summation over indices will always
be assumed unless  explicitly stated. We will also need the convolution integral, which is defined as 
\begin{equation}
(f \otimes g)(z) = \int_0^1 dx dy f(x) g(y) \delta(z - x y)\,. 
\end{equation} 
The kernel $\Gamma_{ij}$ is defined in the $\overline{{\rm MS}}$
 scheme by 
\begin{equation}
\Gamma_{ij}(z) = \delta_{ij} \delta(1-z) 
+ \left( \frac{\alpha_s}{\pi} \right)  \Gamma_{ij}^{(1)}(z)  
+ \left( \frac{\alpha_s}{\pi} \right)^2  \Gamma_{ij}^{(2)}(z)  
+ \mathcal{O}(\alpha_s^3)\, , 
\end{equation}
where the coefficients of the expansion in the strong coupling involve 
the Altarelli-Parisi splitting functions $P^{n}_{ij}$. Specifically, 
\begin{eqnarray}
\Gamma_{ij}^{(1)}(z) &=& -\frac{P^{0}_{ij}(z)}{\epsilon} \, ,\\ 
\Gamma_{ij}^{(2)}(z) &=& -\left\{ \frac{P^{1}_{ij}(z)}{2\epsilon} - \frac{1}{2\epsilon^2}
\left[  
\left(P^{0}_{ik} \otimes P^{0}_{kj}\right)(z)
+ \beta_0 P^{0}_{ij}(z)
\right]
\right\} \, ,
\end{eqnarray}
with $\beta_0 = \frac{11}{4}  - \frac{1}{6} N_F$. 
Let us define the inverse  of the kernel $\Gamma_{ij}$ as 
\begin{equation}
\Delta_{ij}(z) = \sum_{n=0}^2 \Delta_{ij}^{(n)}(z) \left(\frac{\alpha_s}{\pi}\right)^n+\mathcal{O}\left(\alpha_s^3\right) \, ,
\end{equation}
such that it satisfies the condition
 $\left(\Gamma_{ik}\otimes\Delta_{kj}\right)(z) = \delta_{ij}\delta(1-z)$. 
Solving for the coefficients yields
\begin{eqnarray}
\Delta_{ij}^{(0)}(z) &=& \delta_{ij}\delta(1-z)  \, ,\\ 
\Delta_{ij}^{(1)}(z) &=&  -\Gamma_{ij}^{(1)}(z)  =\frac{P^{0}_{ij}(z)}{\epsilon}  \, ,\\ 
\Delta_{ij}^{(2)}(z) &=&  -\Gamma_{ij}^{(2)}(z)  + \left( \Gamma_{ik}^{(1)} \otimes \Gamma_{kj}^{(1)}\right) (z) \nonumber\\
 &=& \frac{P^{1}_{ij}(z)}{2\epsilon} + \frac{1}{2\epsilon^2}  
\left[  
\left(P^{0}_{ik} \otimes P^{0}_{kj}\right)(z)
- \beta_0 P^{0}_{ij}(z)
\right].
\end{eqnarray}
The strong coupling expansion of the bare PDFs then reads  
\begin{equation}
f_{i}(z) = \sum_{n=0}^2 f_{i}^{(n)}(z) \left(\frac{\alpha_s}{\pi}\right)^n+\mathcal{O}\left(\alpha_s^3\right) ,
\label{eq:bare_pdfs}
\end{equation}
with 
\begin{equation}
f^{(n)}_i = \Delta_{ij}^{(n)} \otimes  \tilde{f}_j\,.
\label{eq:kernels} 
\end{equation} 
In evaluating the collinear counter terms 
we encounter convolutions of the type $(f \otimes \Delta )(x)$, where  the function $f$ is  regular 
and $\Delta (x)$ can in general be written as 
\begin{equation}
\Delta (x) = a \delta(1-x) + \sum_n b_n\Di{n}(x) + C(x).
\end{equation}
Expressing the convolution as a single integral we obtain
\begin{equation}
(f \otimes \Delta )(x)  = 
 \int_x^1  \frac{dy}{y} \, f\left( \frac{x}{y}\right) 
\left\{ a \delta(1-y) + \sum_n b_n\Di{n}(y) + C(y)  \right\} .
\end{equation}
Care has to be taken with convolutions over $\Di{n}$. Since the integration does not start at zero, a boundary term must be included
\begin{equation}
\left( \Di{m} \otimes f \right)(x) = \frac{\log(1-x)^{m+1}}{m+1} f(x)
+ \int_{x}^{1} dy \,\log(1-y)^{m} \frac{\frac{1}{y}  f\left(\frac{x}{y}\right)-f(x)}{1-y}\,.
\end{equation}
Because of the downward sloping shape of all parton distribution functions, a quadratic remapping of the integration variable
$y$ was found to optimize the convergence behavior, i.e. we parametrized the integral like
\begin{equation*}
 y = x + (1-x)z^2 ,
\end{equation*}
with $z$ uniformly distributed between $0$ and $1$.

In our code, this integration is carried out numerically. The integration is  one-dimensional, which makes a simple
deterministic trapezium integration with about $50.000$ points the simplest option. The result of the integration is 
accurate to at least 5 digits, which is usually below the precision of the  Monte Carlo integration. The precision of the integration can be arbitrarily increased by increasing the number of points used. For every bare PDF used, we construct a one-dimensional grid in the Bjorken-$x$ variable and interpolate from it during runtime. 

An alternative to constructing a grid is to perform the integration numerically along with the phase space ones, thereby increasing the dimensionality of the Monte Carlo integration by one (or by two in the case of double NLO kernels convoluted with the Born). We have implemented this as well and found that it yields the same results as the grid approach.

This procedure allows us to expand the (singular) bare PDFs via eq.(\ref{eq:bare_pdfs}) order by order in the dimensional regulator $\epsilon$ and substitute them directly in eq.(\ref{master}). The singularities in the resulting convolutions, appearing as poles in the $\epsilon$-expansion, cancel the initial state collinear singularities of the partonic cross section. This cancellation is achieved numerically in our calculation and can be observed bin by bin in e.g. the rapidity distribution of the Higgs boson. One can achieve this cancellation in each initial state channel separately, at the cost of separating the convolution integrals depending on the initial state parton in the convolution, i.e. by not performing the implicit $j$-summation in eq.(\ref{eq:kernels}).

It is worth pointing out that the procedure described here is entirely generic, i.e. it provides the collinear counter terms for any NNLO process numerically. Moreover, we thereby circumvent the usual insertion of eq.(\ref{eq:renormed_pdf}) in the equivalent of eq.(\ref{master}) for renormalized quantities and the resulting cumbersome and process specific analytic treatment of the convolutions.


\section{Numerical results}
\label{sec:numerical_results}
 We have performed a number of tests to ensure that our results  are consistent with each other and with results  available in the literature:
\begin{itemize}
\item We have implemented the entire calculation in two different computer codes, one in {\tt Fortran} and one in {\tt C++}, and all results agree within their respective Monte Carlo errors, both inclusively and differentially.
\item The coefficients of all poles in the $\epsilon$-expansion of all cross sections cancel both inclusively and differentially for the entire process and also for all individual initial state channels.
\item The inclusive cross section agrees with the one available in~\cite{Harlander:2003ai} and from {\tt ihixs}~\cite{Anastasiou:2011pi} and so does the inclusive cross section per initial state channel. This is the first independent check of the inclusive cross section published in~\cite{Harlander:2003ai} and adopted in~\cite{Anastasiou:2011pi}.
\item The soft limit of both real-virtual and double real contributions were computed both numerically (as a limiting case of the generic matrix elements) and analytically. Moreover the integrated double real contributions were found to agree with an analytic computation provided by~\cite{Anastasiou:private}.
\item The subtraction process for every double real integral was implemented in two different ways and were found in complete agreement.
\end{itemize}


\begin{figure}[htbp]
\centering
\includegraphics[width=0.8\textwidth]{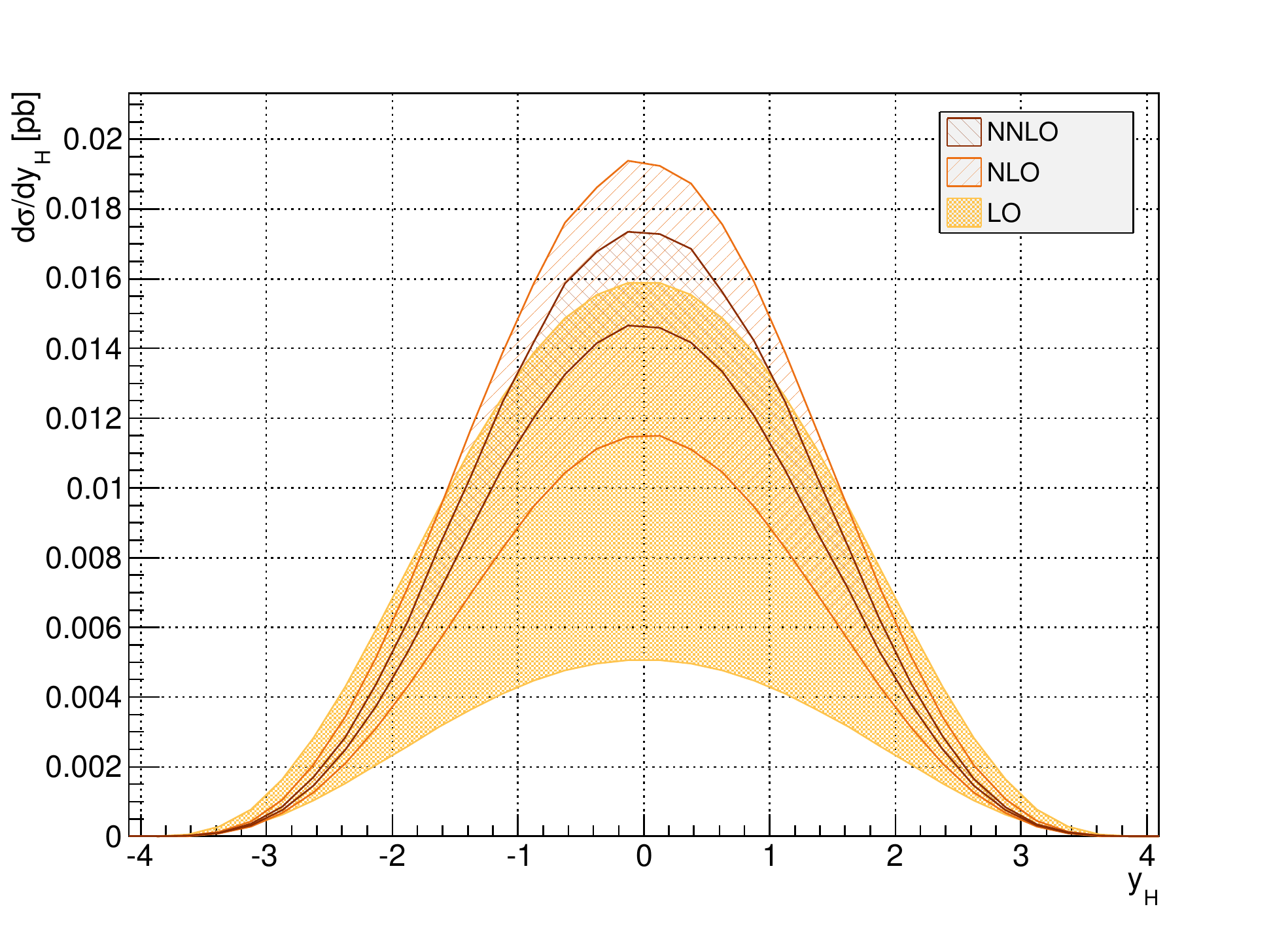}
\caption{The Higgs rapidity distribution for $m_H = 125$ GeV at the $8$ TeV LHC. The bands describe the uncertainty due to factorization scale}
\label{plot:higgs_rapidity}
\end{figure}

\begin{figure}[htbp]
\centering
\includegraphics[width=0.8\textwidth]{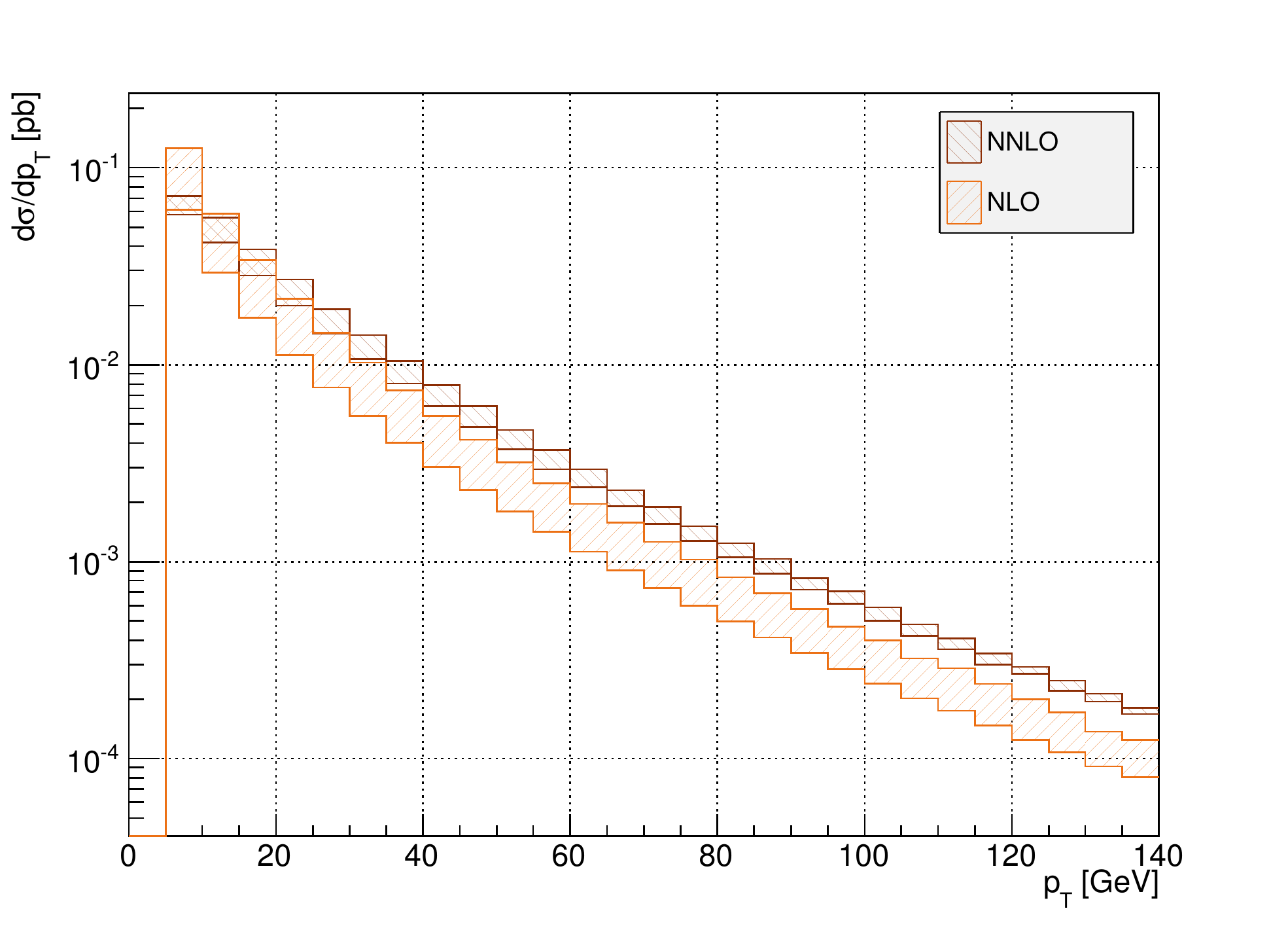}
\caption{The Higgs transverse momentum distribution for $m_H = 125$ GeV at the $8$ TeV LHC.}
\label{plot:higgs_PT}
\end{figure}

We present results for the LHC with a center of mass energy of $8$ TeV. We fix the mass of the Higgs boson at $125$ GeV. We have used the MSTW2008 ($68\%$CL) PDFs for all results presented here. The value of $\alpha_s$ at $m_Z$ that we use is the best-fit value of the PDF set at the corresponding order. We use $\mu_R=m_H$ as the central renormalization scale. The value of $\alpha_s$ used is run from $m_Z$ to $\mu_R$ through NNLO in QCD. The mass of the bottom quarks is set to zero in all matrix elements,  consistently with the 5FS choice. The bottom Yukawa coupling, however, depends on the mass of the bottom. The Yukawa coupling at $\mu_R$ is obtained from the Yukawa coupling at $\mu^*=10$ GeV, using $m_b(\mu^*)=3.63$ GeV.  

We do not vary $\mu_R$ in what follows, since the $\mu_R$ scale dependence of the total cross section has been found to be very mild. We have also checked that the $\mu_R$-dependence of differential distributions is very small.

Previous studies have shown that the inclusive cross section is very sensitive to the choice of factorization scale. Arguments related to the validity of the 5FS approximation with respect to the collinearity of final state $b$-quarks, as well as to the matching to the 4FS calculation or to the need for a smoother perturbative expansion, point to factorization scales that are much lower than the Higgs boson mass. We adopt the choice $\mu_F={ m_H\over 4}$ as a central scale and vary it in the range $[{m_H\over 8},{m_H\over 2}]$ to estimate the related uncertainty.

All Monte Carlo integrations was performed with the {\tt Cuba}~\cite{Hahn:2004fe} implementation of the Vegas algorithm. 

The rapidity distribution of the Higgs boson is shown at fig.~\ref{plot:higgs_rapidity}. As expected, the perturbative expansion is converging smoothly for this choice of central $\mu_F$ and the NNLO uncertainty band is entirely engulfed by the NLO one.

The transverse momentum distribution for the Higgs boson is shown in fig.~\ref{plot:higgs_PT}. This observable starts at NLO in QCD in the 5FS, and the fixed order prediction fails, as usual, to describe the very low $p_T$ spectrum due to the related large logarithms. At the large $p_T$ range we see that the NNLO calculation leads to a harder spectrum than the NLO one and the NLO scale uncertainty fails to capture this feature. This implies that great care should be taken when relying on NLO predictions 
for observables that are highly exclusive in the transverse momentum of the Higgs boson.

\begin{figure}[htbp]
\centering
\begin{minipage}[b]{0.49\textwidth}
\includegraphics[width=\textwidth]{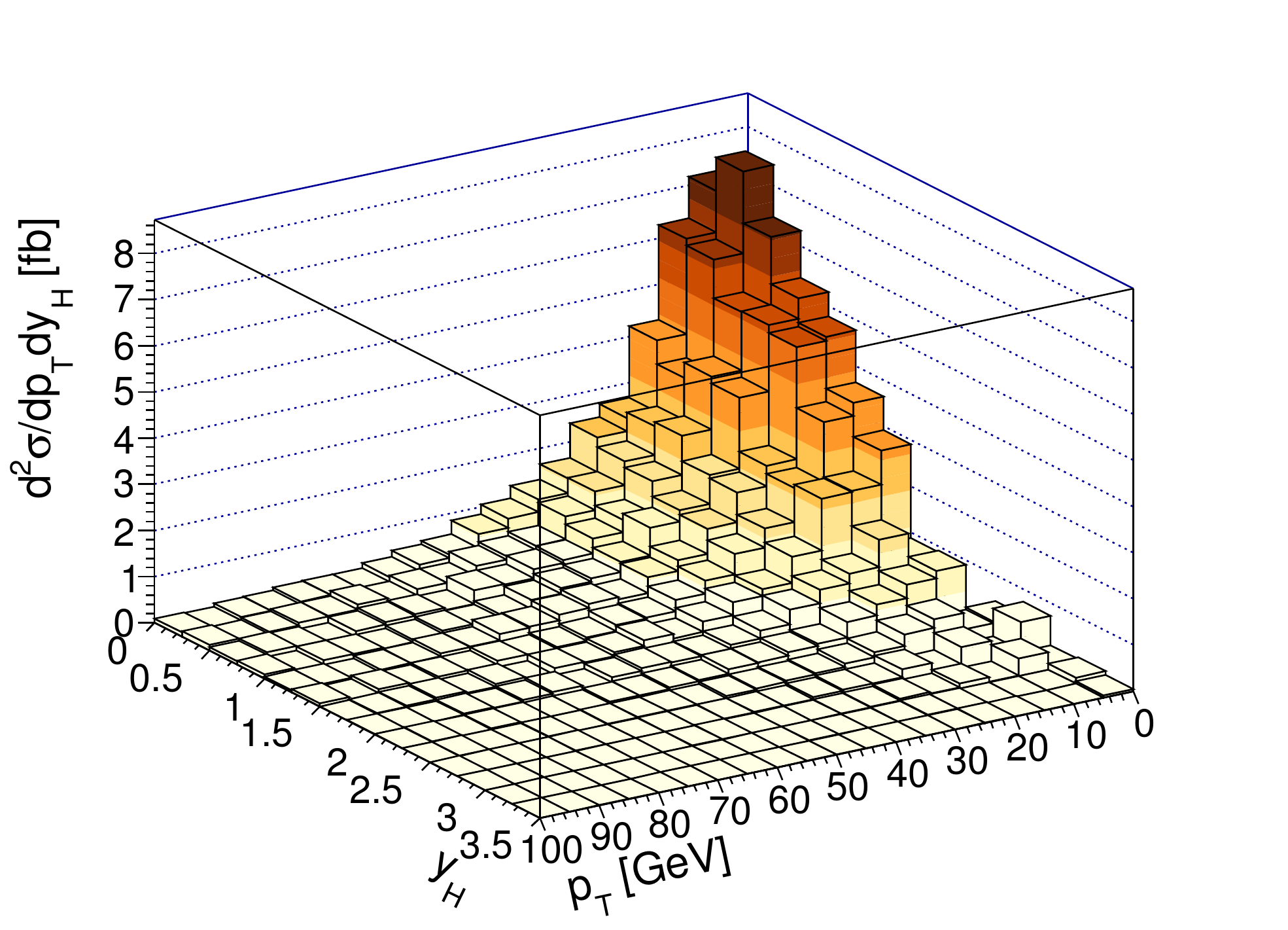}
\end{minipage}
\begin{minipage}[b]{0.49\textwidth}
\includegraphics[width=\textwidth]{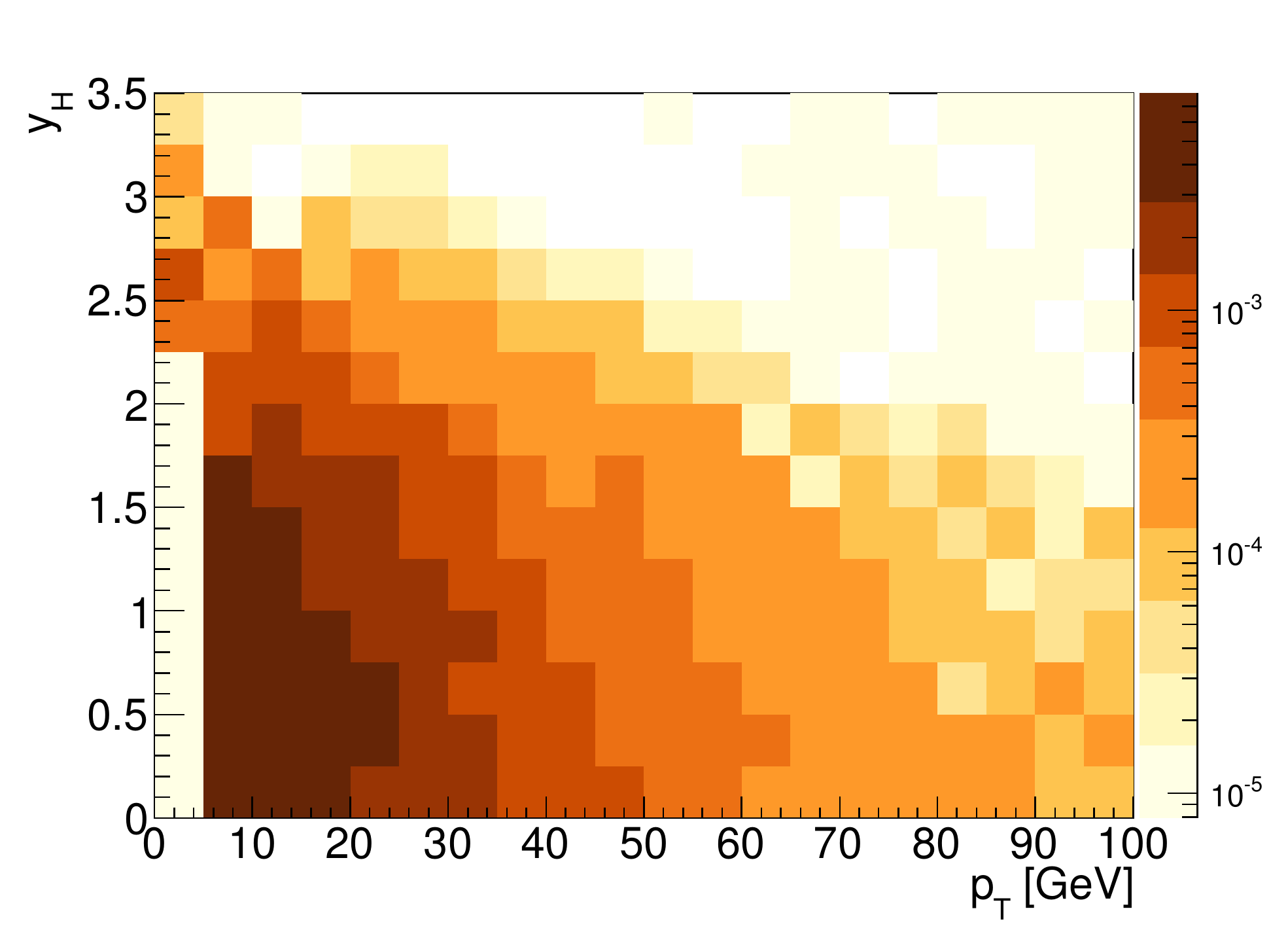}
\end{minipage}
\caption{Differential distribution in rapidity and transverse momentum of the Higgs boson for $m_H = 125$ GeV at the $8$ TeV LHC. }
\label{plot:2d_hist}
\end{figure}

The differential distribution in both the rapidity and the $p_T$ of the Higgs is shown in fig.~\ref{plot:2d_hist}, both in a three-dimensional lego plot and in a density plot. We see that the bulk of the events are produced centrally (with $|y|<2.5$) and at relatively low $p_T$ ($~35-50$GeV).  

\begin{figure}[htbp]
\centering
\includegraphics[width=0.8\textwidth]{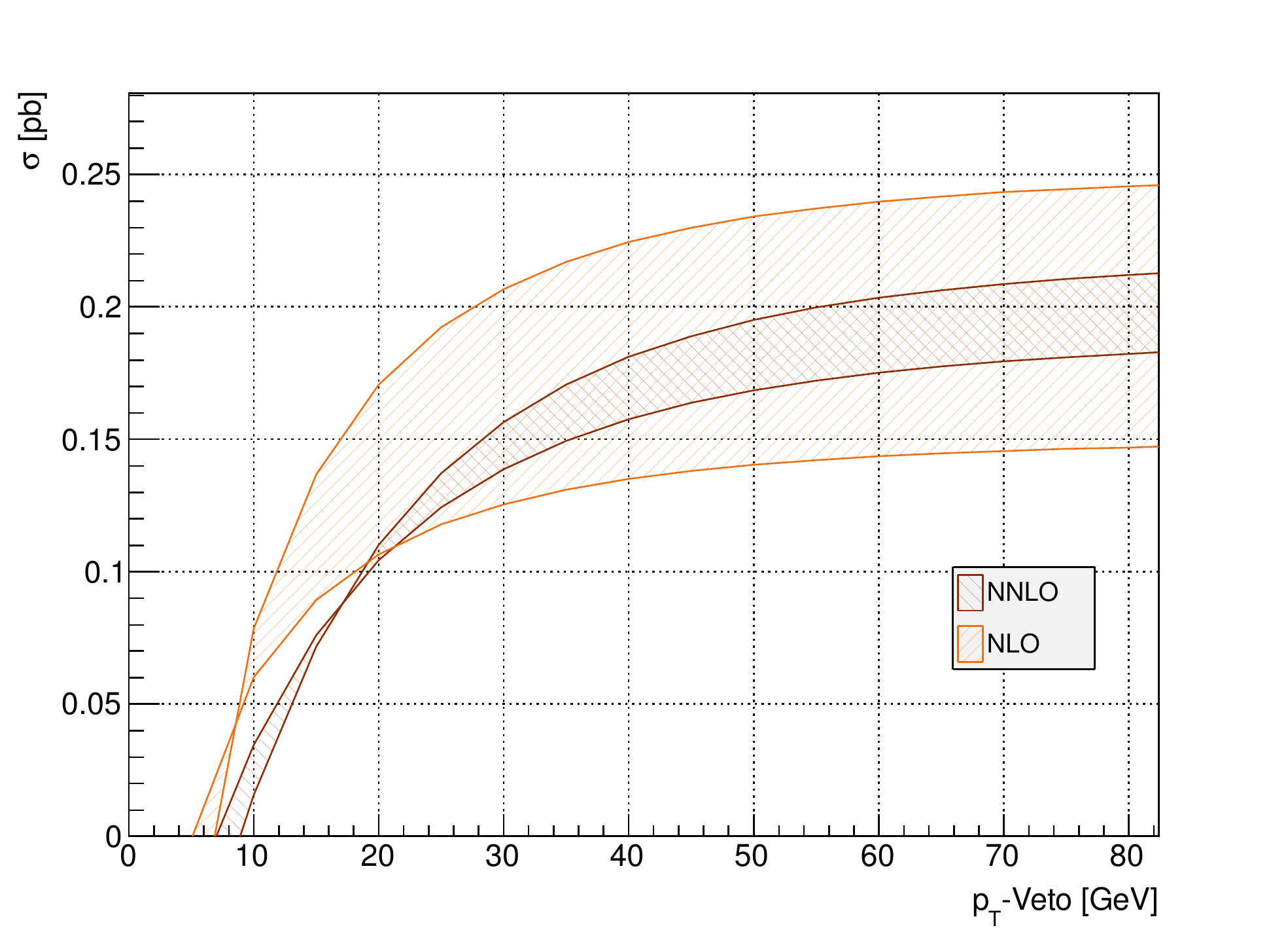}
\caption{The cumulative distribution of the Higgs  $p_T$ for $m_H = 125$ GeV at the $8$ TeV LHC.}
\label{plot:higgs_pt_veto}
\end{figure}
\begin{figure}[htbp]
\centering
\includegraphics[width=0.8\textwidth]{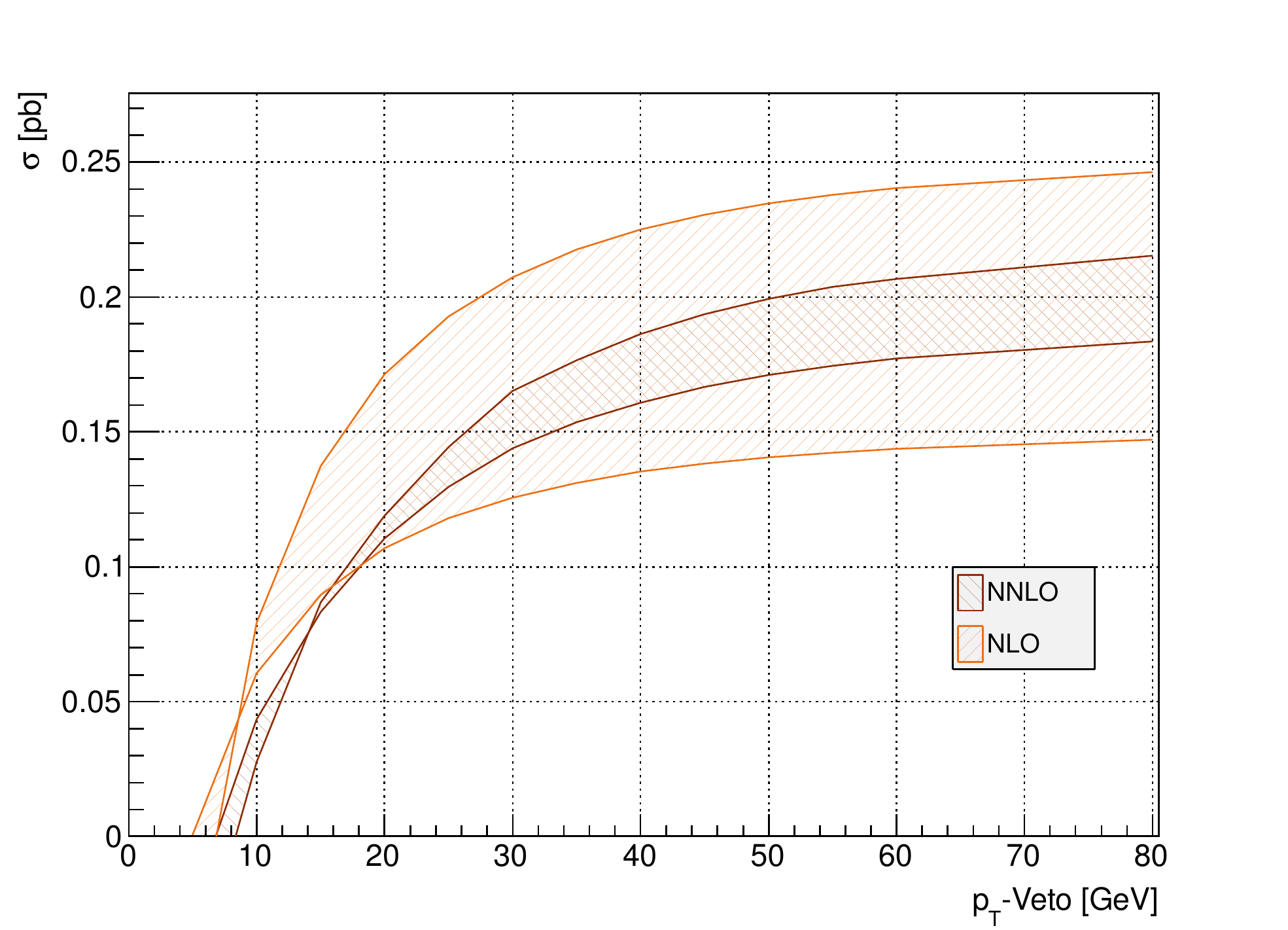}
\caption{The cross section in the presence of a jet veto (the $0$-jet rate)  for $m_H = 125$ GeV at the $8$ TeV LHC.}
\label{plot:jet_veto}
\end{figure}

In fig.~\ref{plot:higgs_pt_veto} we show the cumulative distribution of the Higgs transverse momentum. This observable is equivalent to the cross section in the presence of a jet veto at NLO, but only related indirectly at NNLO. In fig.~\ref{plot:jet_veto} we present the cross section in the presence of a jet veto. We see again that the perturbative description for high $p_T$ cut-offs is satisfactory (despite the discrepancy in high $p_T$ between NLO and NNLO, which is, in absolute terms, unimportant), while for cut-offs lower than $20$ GeV the NLO description does not coincide with the NNLO one. The vanishing of the uncertainty around $15$ GeV (which in the case of the jet veto is taking place at a slightly lower $p_T$-veto value) is a feature reminiscent of a similar situation in Higgs production via gluon fusion \cite{Anastasiou:2008ik}. The fixed order prediction in this region is very stable under varying the factorization scale, and any residual uncertainty in quantities like the acceptance in the presence of a veto is driven by the uncertainty in the total cross section. Various approaches to assign a larger uncertainty to similar observables involving re-summation exist, see for example \cite{Banfi:2012yh}.

\begin{figure}[htbp]
\centering
\begin{minipage}[b]{0.49\textwidth}
\includegraphics[width=\textwidth]{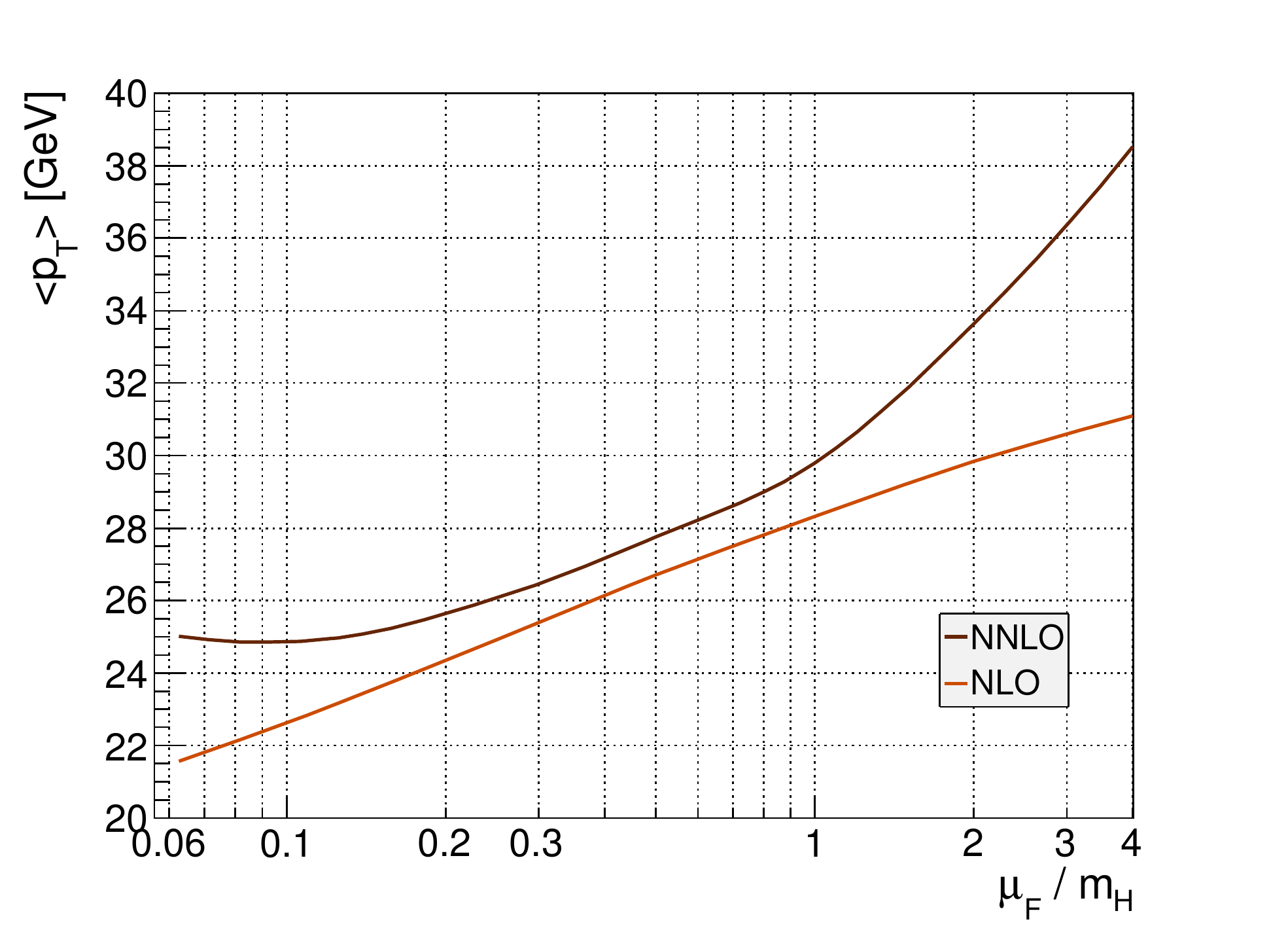}
\end{minipage}
\begin{minipage}[b]{0.49\textwidth}
\includegraphics[width=\textwidth]{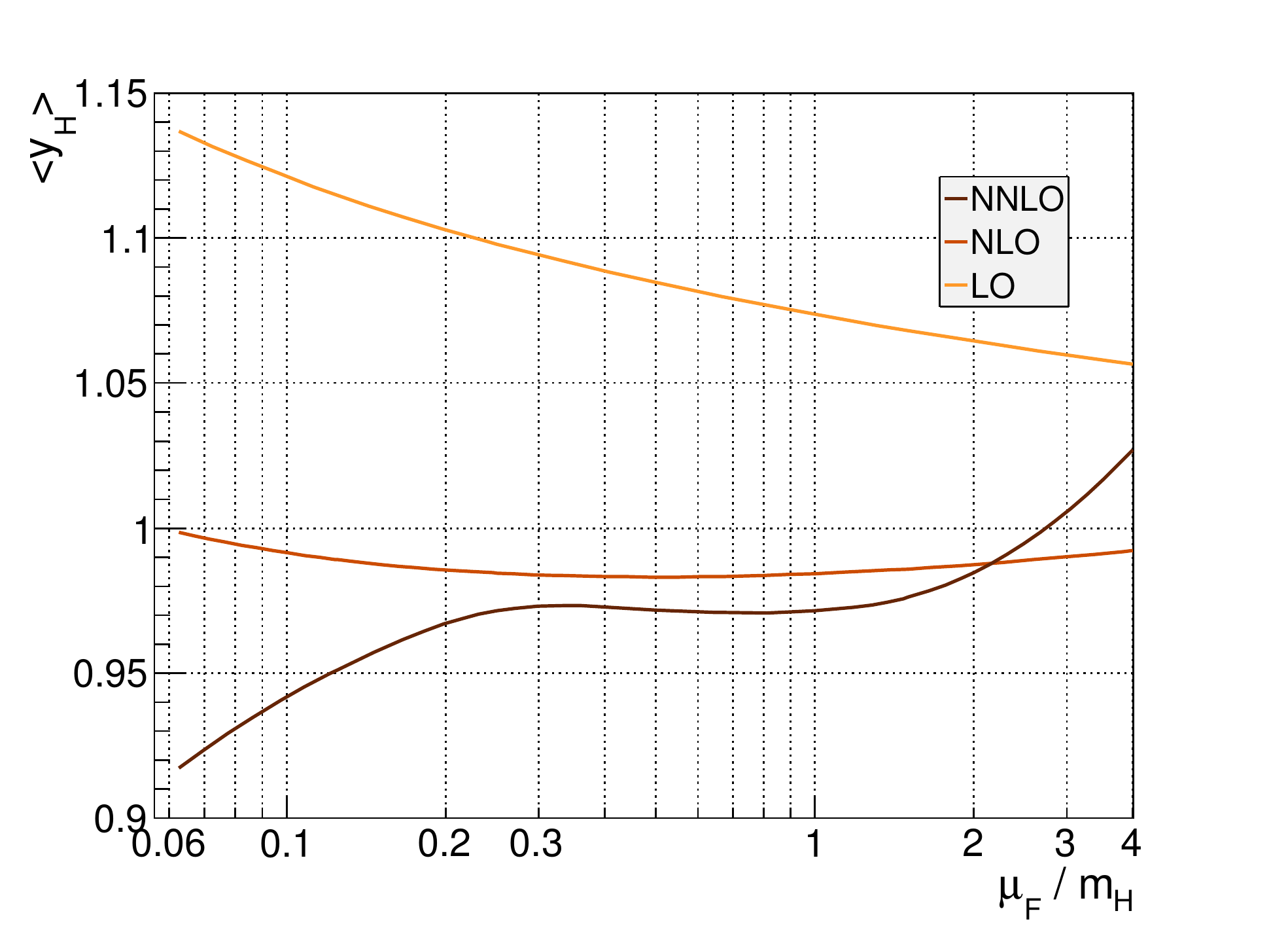}
\end{minipage}
\caption{The average $p_T$ (left) and rapidity (right) of the Higgs as a function of $\mu_F/m_H$ for $m_H = 125$ GeV at the $8$ TeV LHC.}
\label{plot:av_pt_rap}
\end{figure}

\begin{figure}[htbp]
\centering
\includegraphics[width=0.8\textwidth]{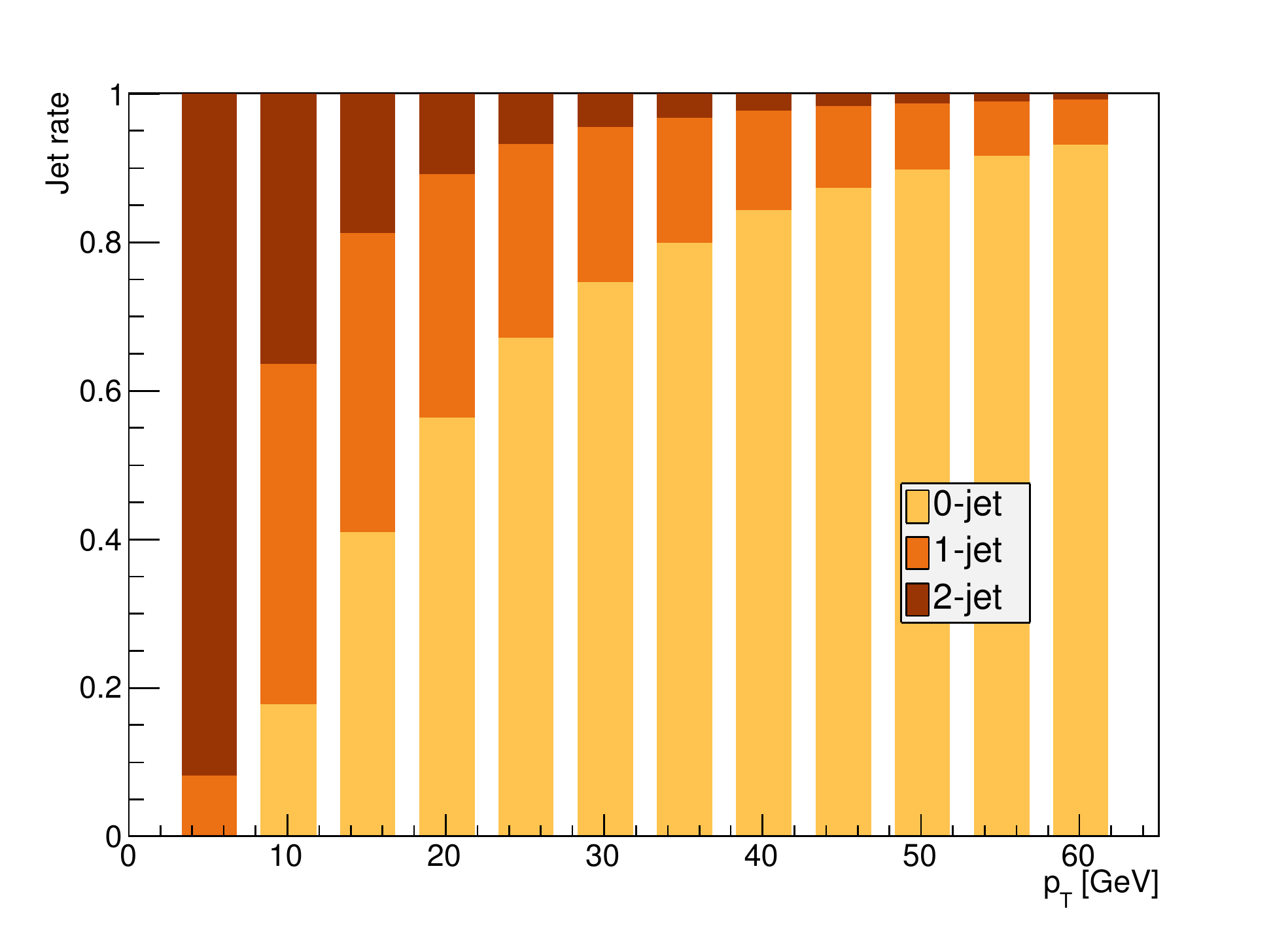}
\caption{The $0$-, $1$- and $2$-jet rate as a function of the $p_T$ used in the jet definition for $m_H = 125$ GeV at the $8$ TeV LHC.}
\label{plot:jet_rates}
\end{figure}

An important observable in $b\bar{b}\to H$ is the cross section for zero, one and two jets.  We use the anti-$k_T$ algorithm~\cite{Cacciari:2008gp} for jet clustering\footnote{At this order in perturbation theory, the anti-$k_T$, the $k_T$ and the Cambridge-Aachen algorithms are completely equivalent.} with a cone in the $y-\phi$ plane of radius  $R=0.4$. We show in fig.~\ref{plot:jet_rates} the jet rates as a function of the jet $p_T^{max}$ used to define them. Here we do not distinguish between $b$-jets and light jets. We find the jet rates for $p_{T}^{max}=20$GeV to be in agreement with those published 
in~\cite{Harlander:2011fx}.

\begin{figure}[htbp]
\centering
\begin{minipage}[b]{0.49\textwidth}
\includegraphics[width=\textwidth]{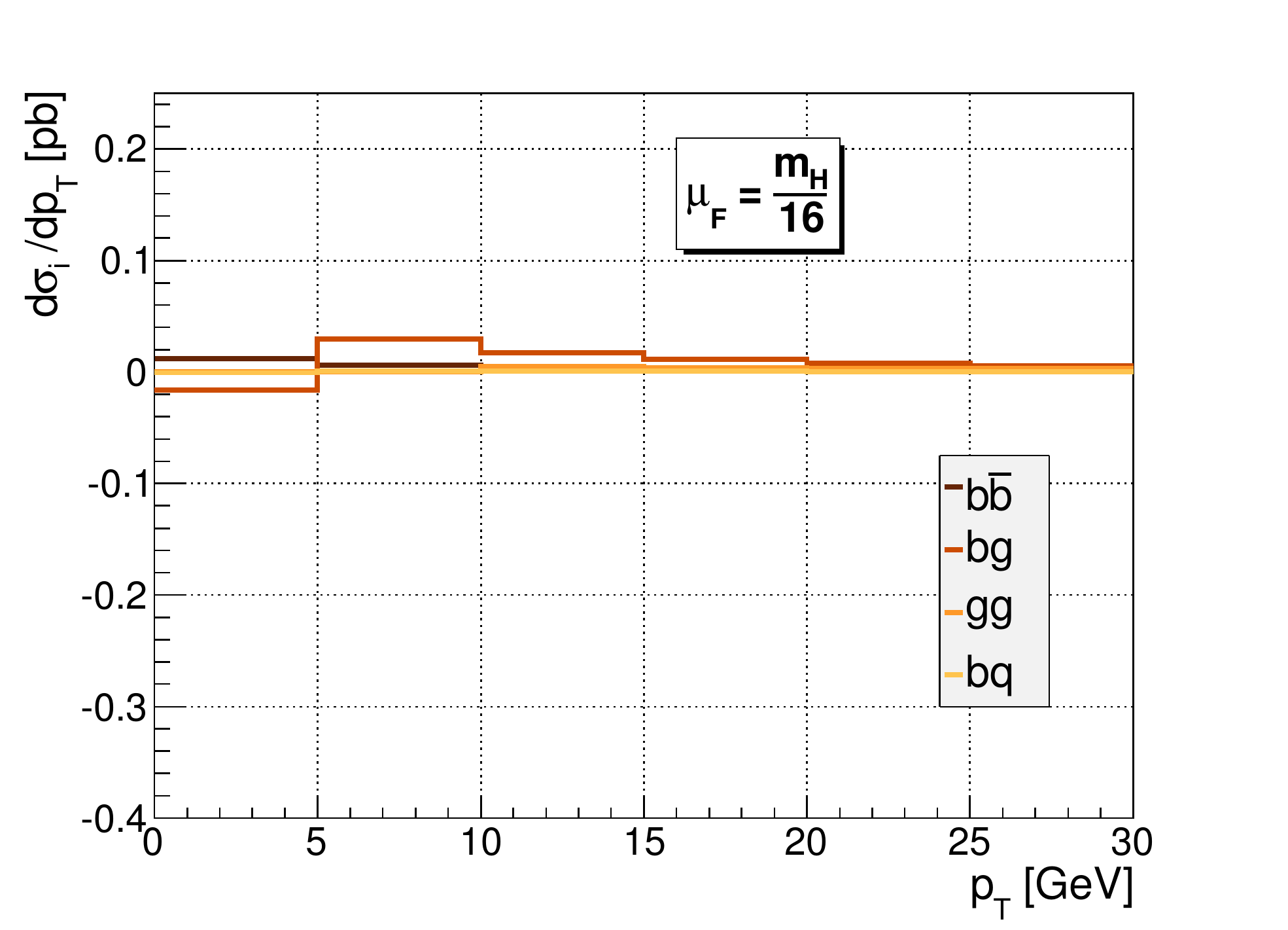}
\end{minipage}
\begin{minipage}[b]{0.49\textwidth}
\includegraphics[width=\textwidth]{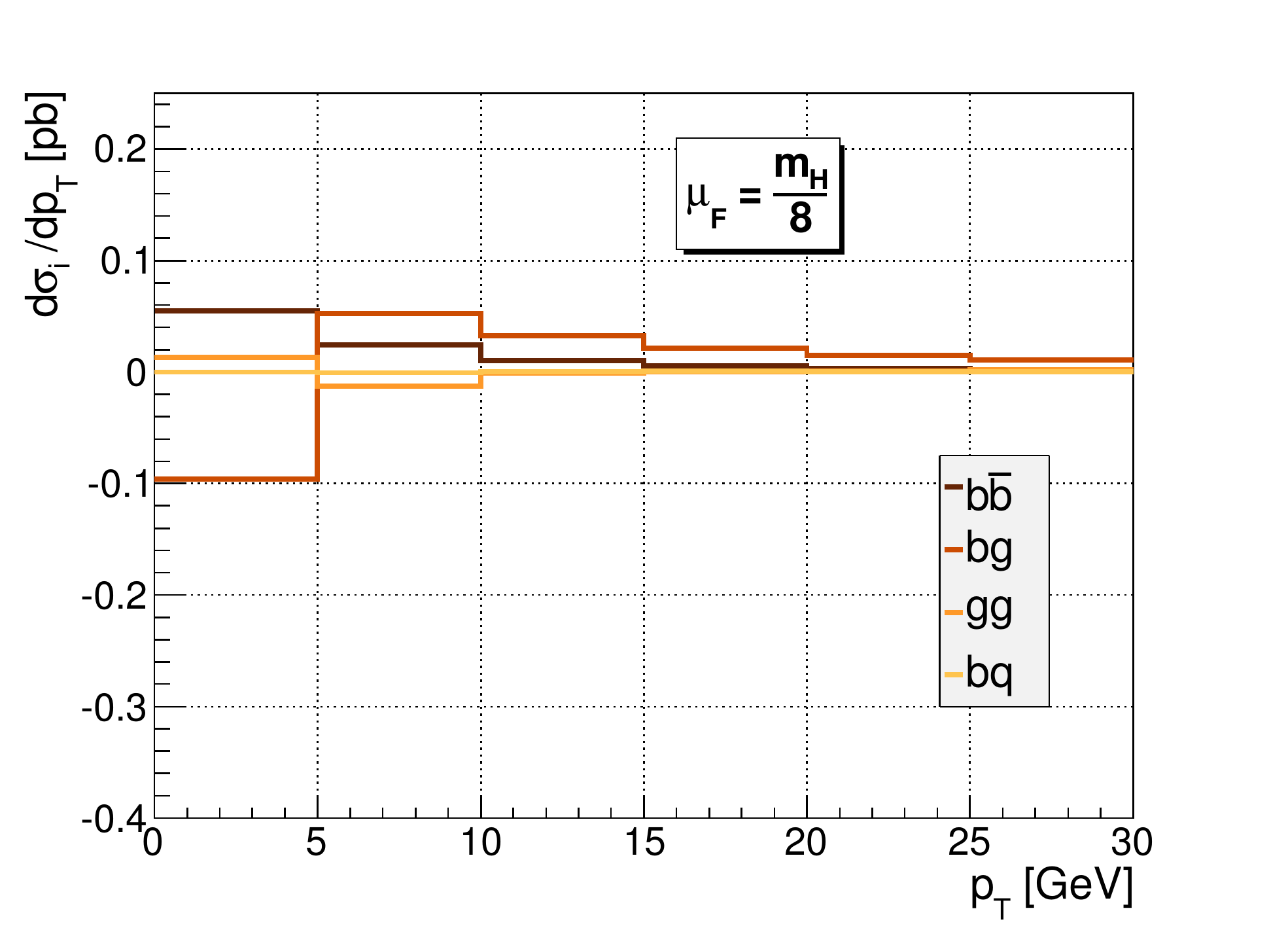}
\end{minipage}
\begin{minipage}[b]{0.49\textwidth}
\includegraphics[width=\textwidth]{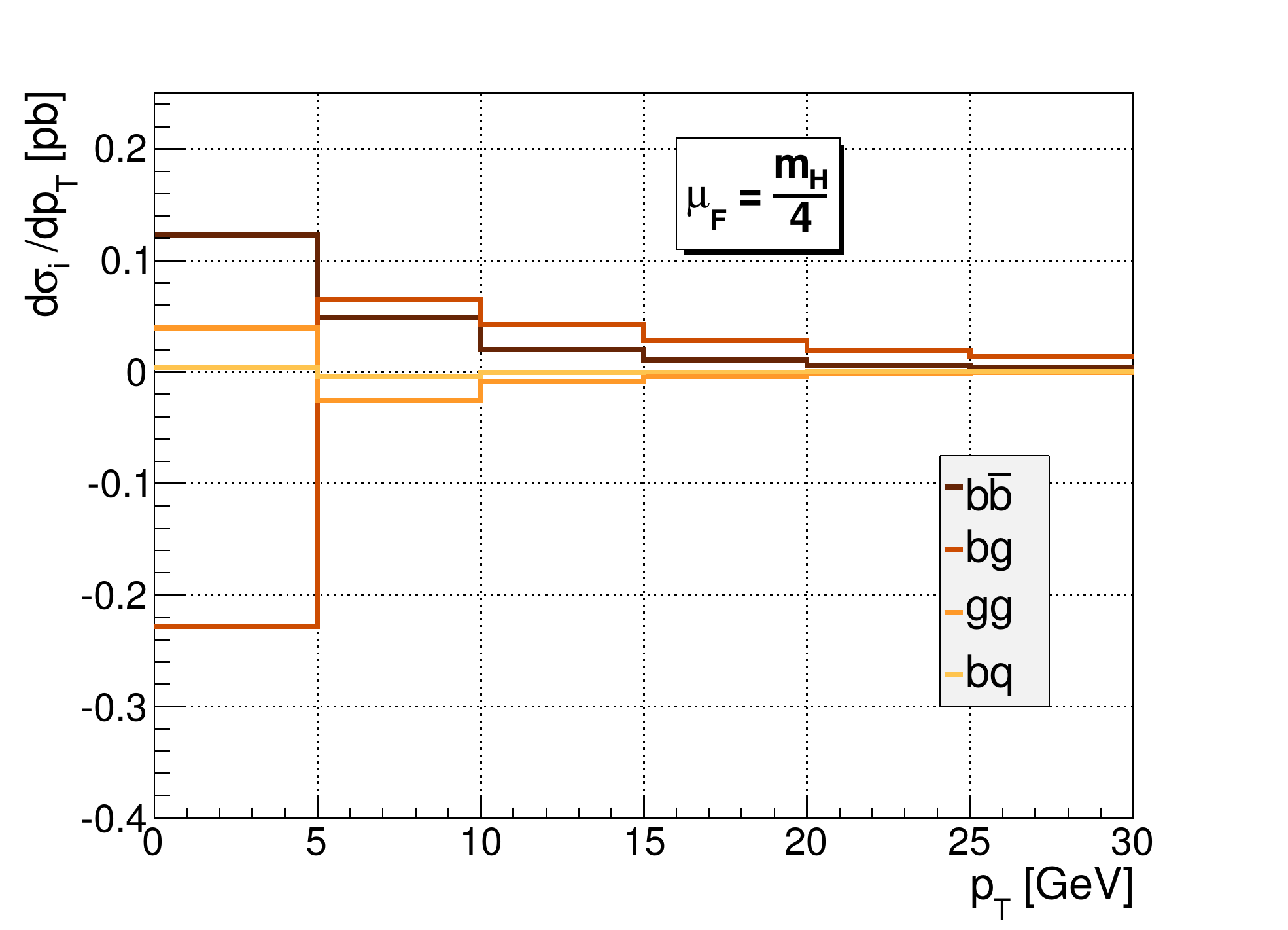}
\end{minipage}
\begin{minipage}[b]{0.49\textwidth}
\includegraphics[width=\textwidth]{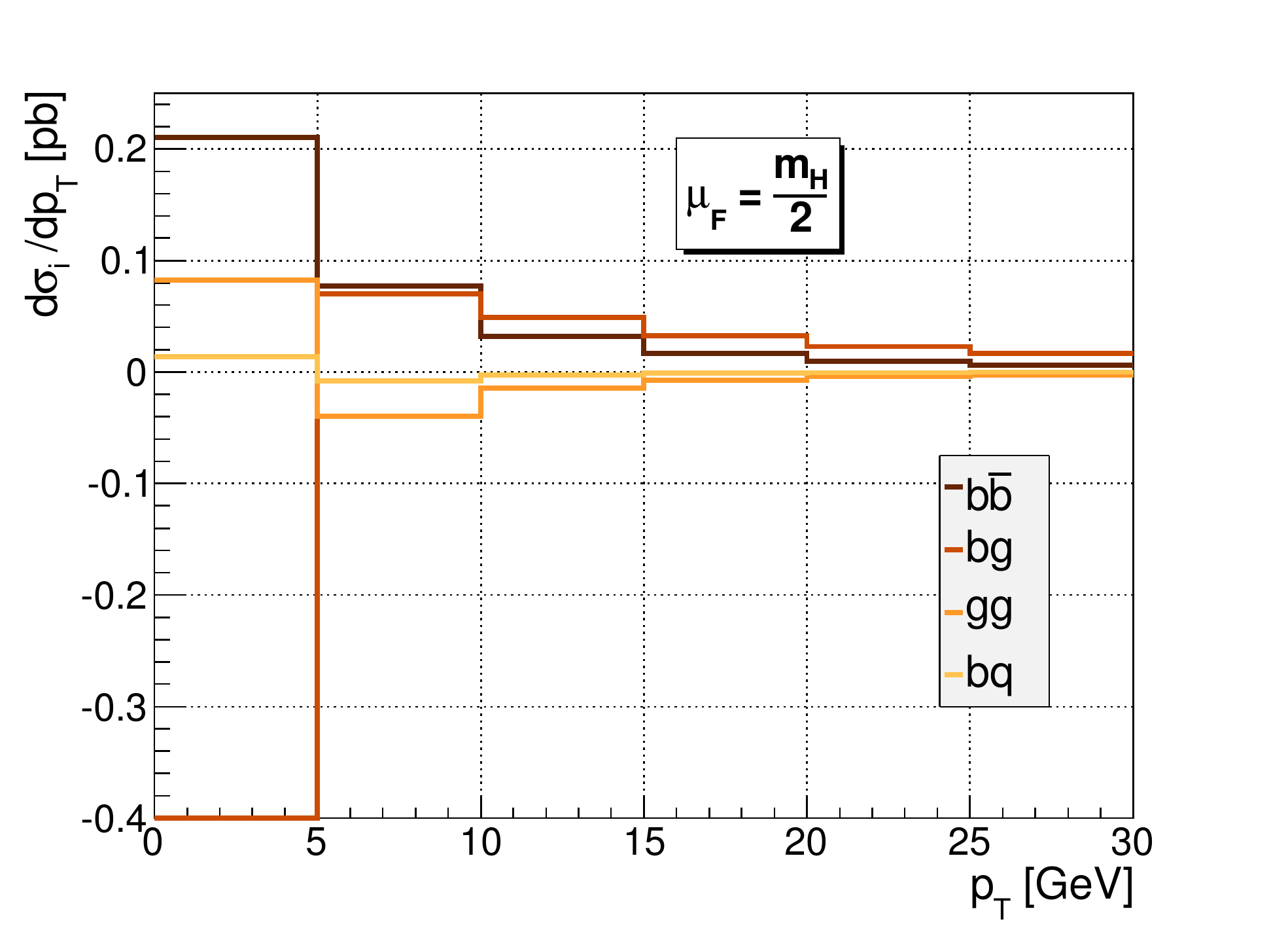}
\end{minipage}
\begin{minipage}[b]{0.49\textwidth}
\includegraphics[width=\textwidth]{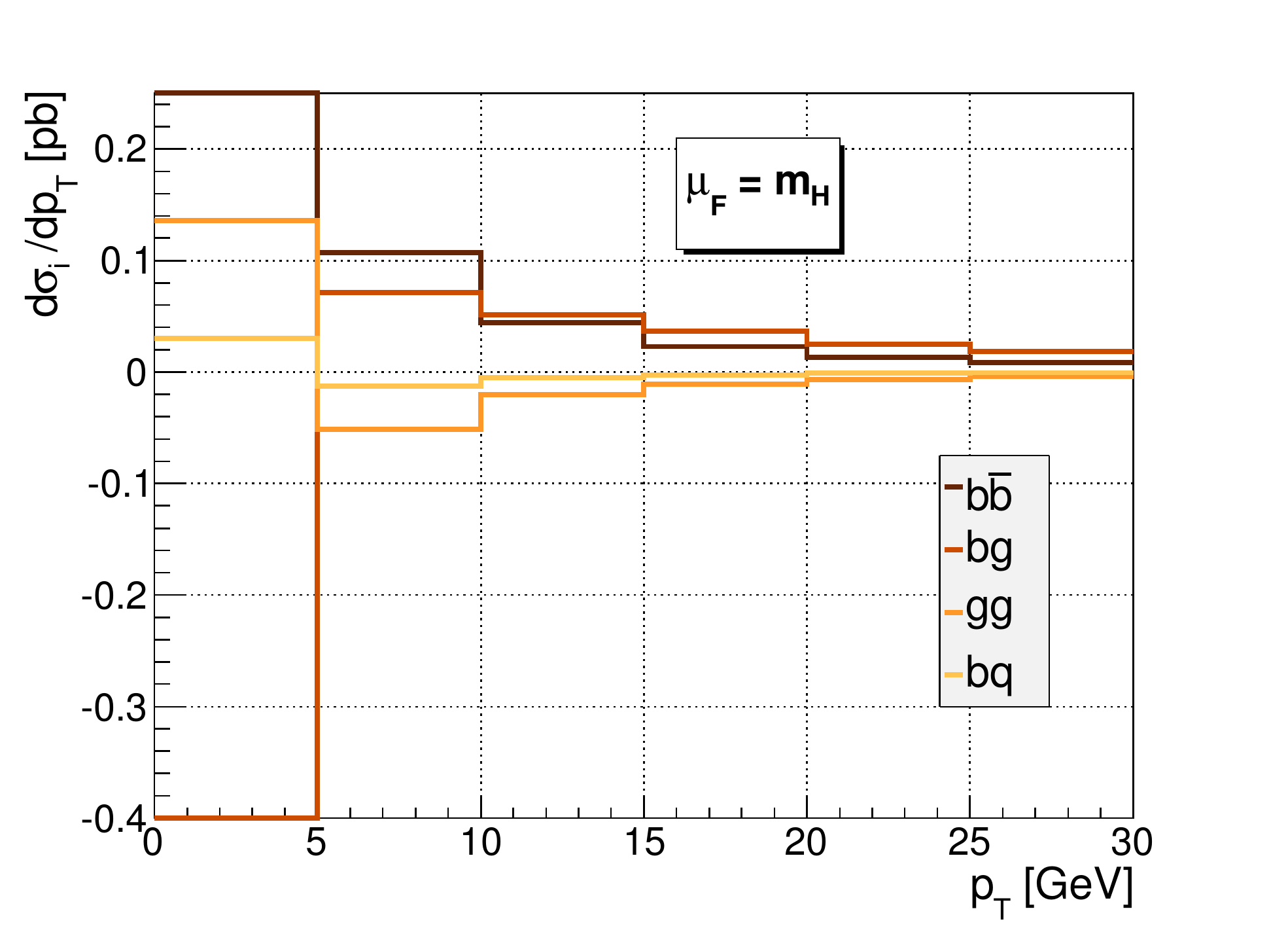}
\end{minipage}
\begin{minipage}[b]{0.49\textwidth}
\includegraphics[width=\textwidth]{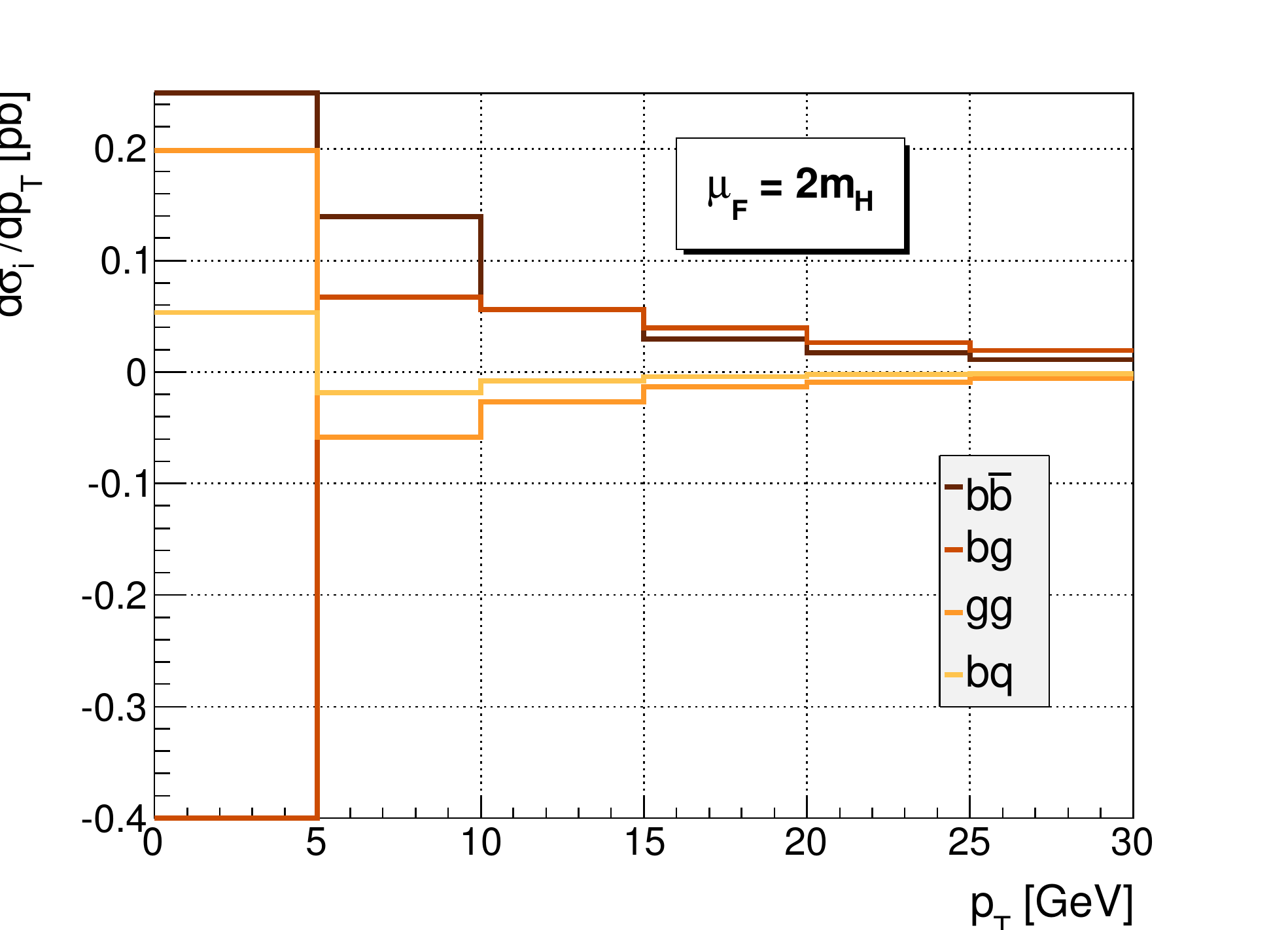}
\end{minipage}
\caption{The distribution of the Higgs $pT$ per initial state channel for $m_H = 125$ GeV at the $8$ TeV LHC, with 
$\mu_F={m_H\over 16},{m_H\over 8},{m_H\over 4},{m_H\over 2},m_H,2m_H$.}
\label{plot:channels_PT}
\end{figure}

\begin{figure}[htbp]
\centering
\begin{minipage}[b]{0.49\textwidth}
\includegraphics[width=\textwidth]{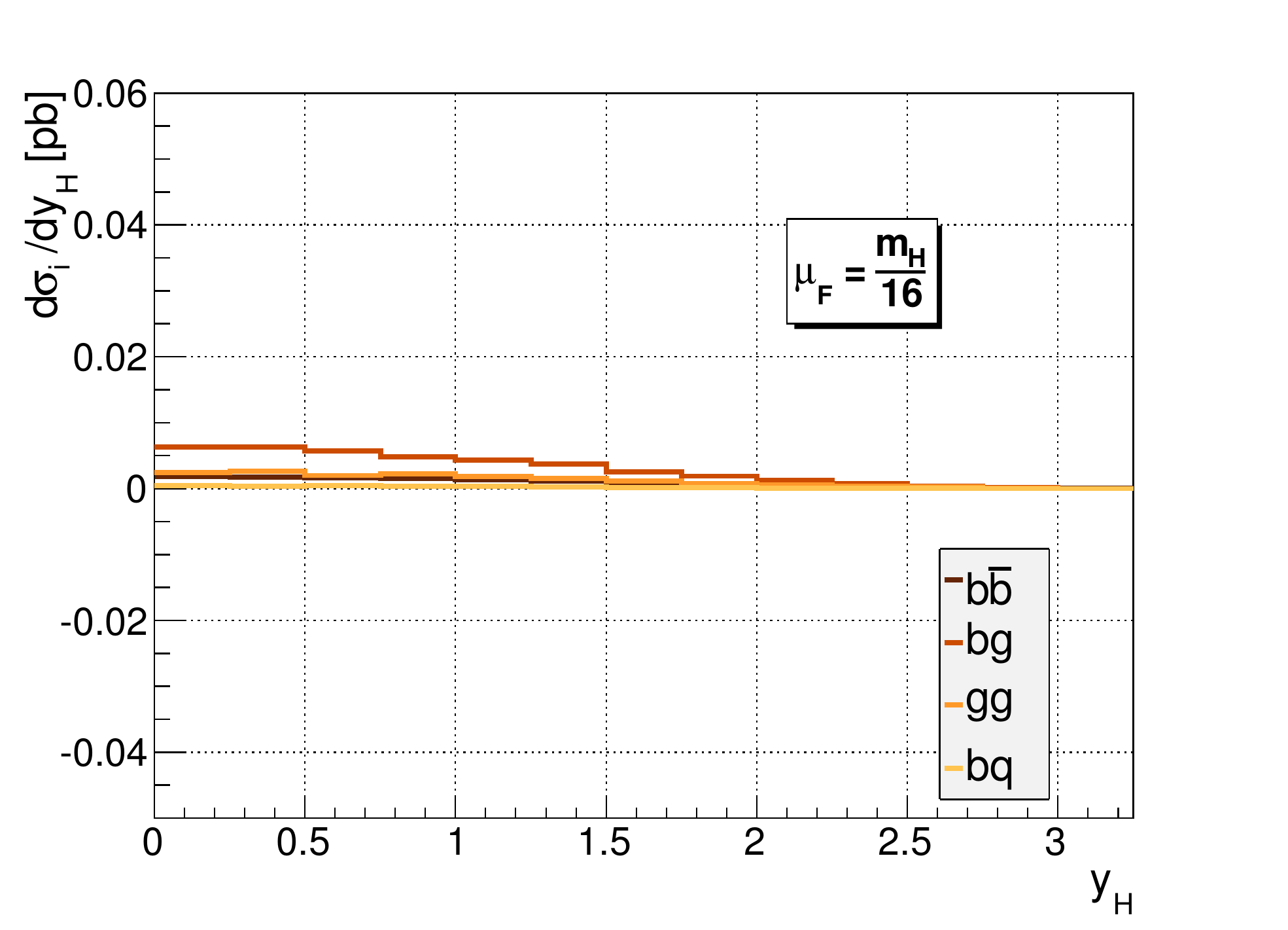}
\end{minipage}
\begin{minipage}[b]{0.49\textwidth}
\includegraphics[width=\textwidth]{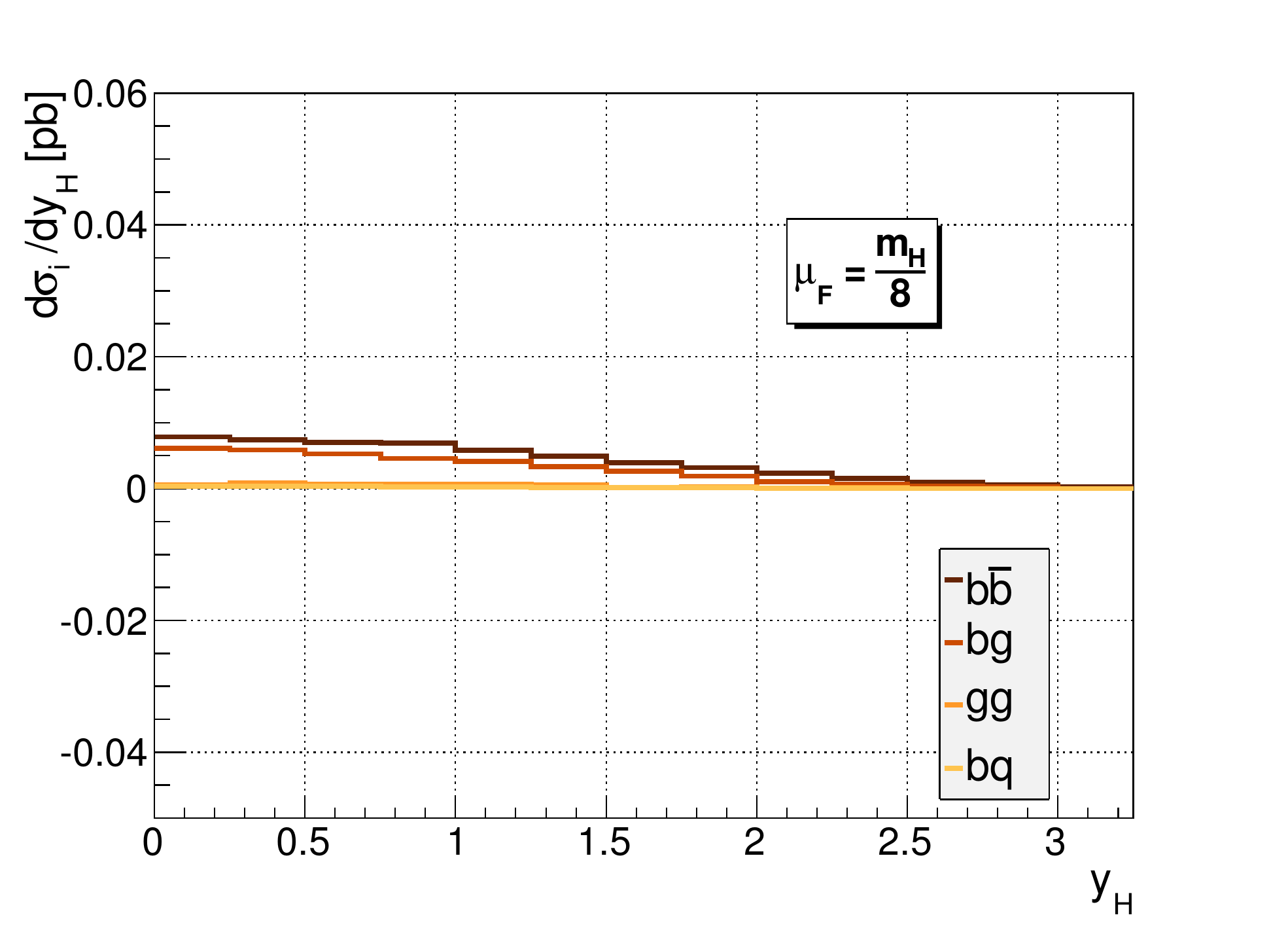}
\end{minipage}
\begin{minipage}[b]{0.49\textwidth}
\includegraphics[width=\textwidth]{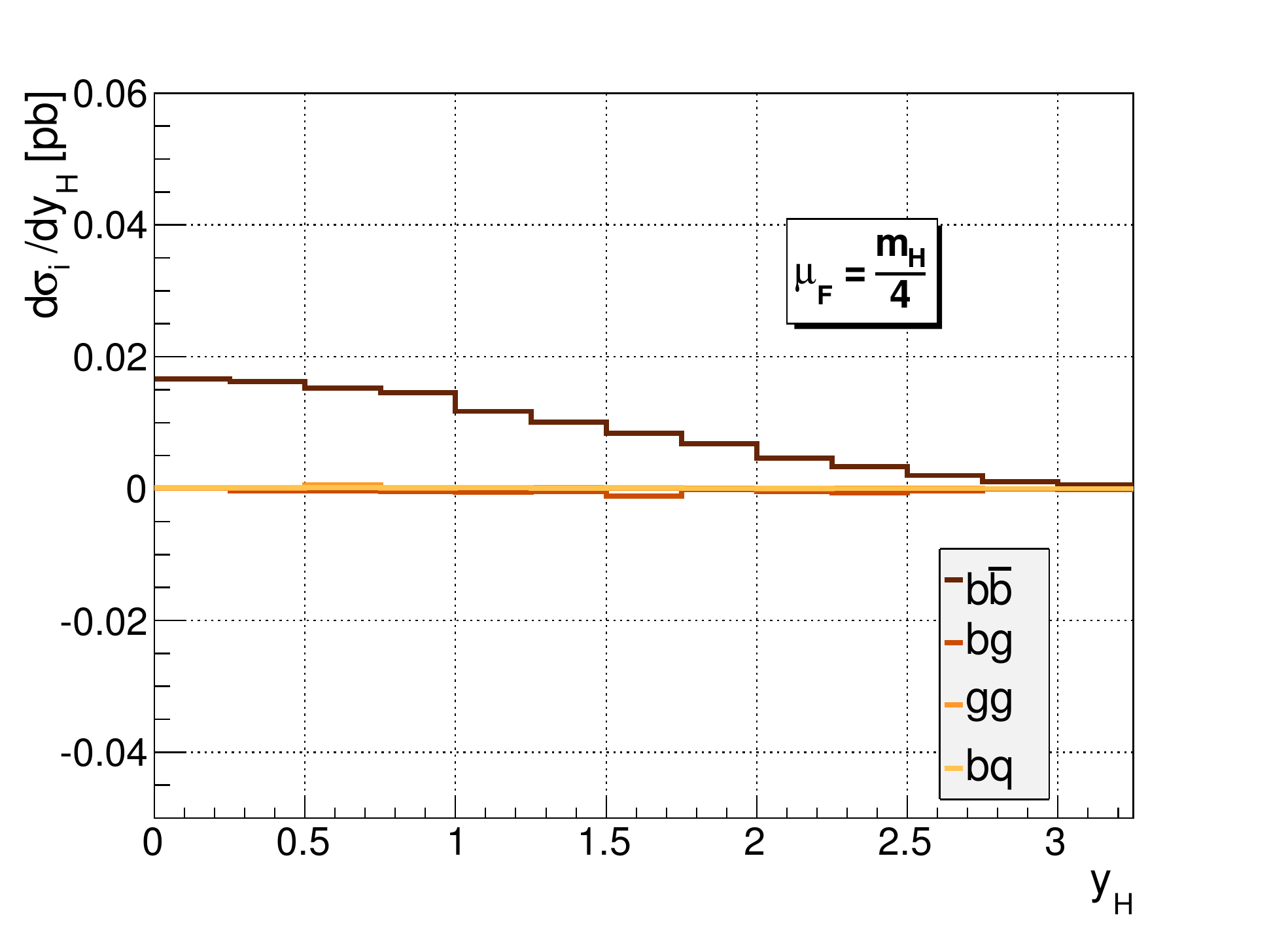}
\end{minipage}
\begin{minipage}[b]{0.49\textwidth}
\includegraphics[width=\textwidth]{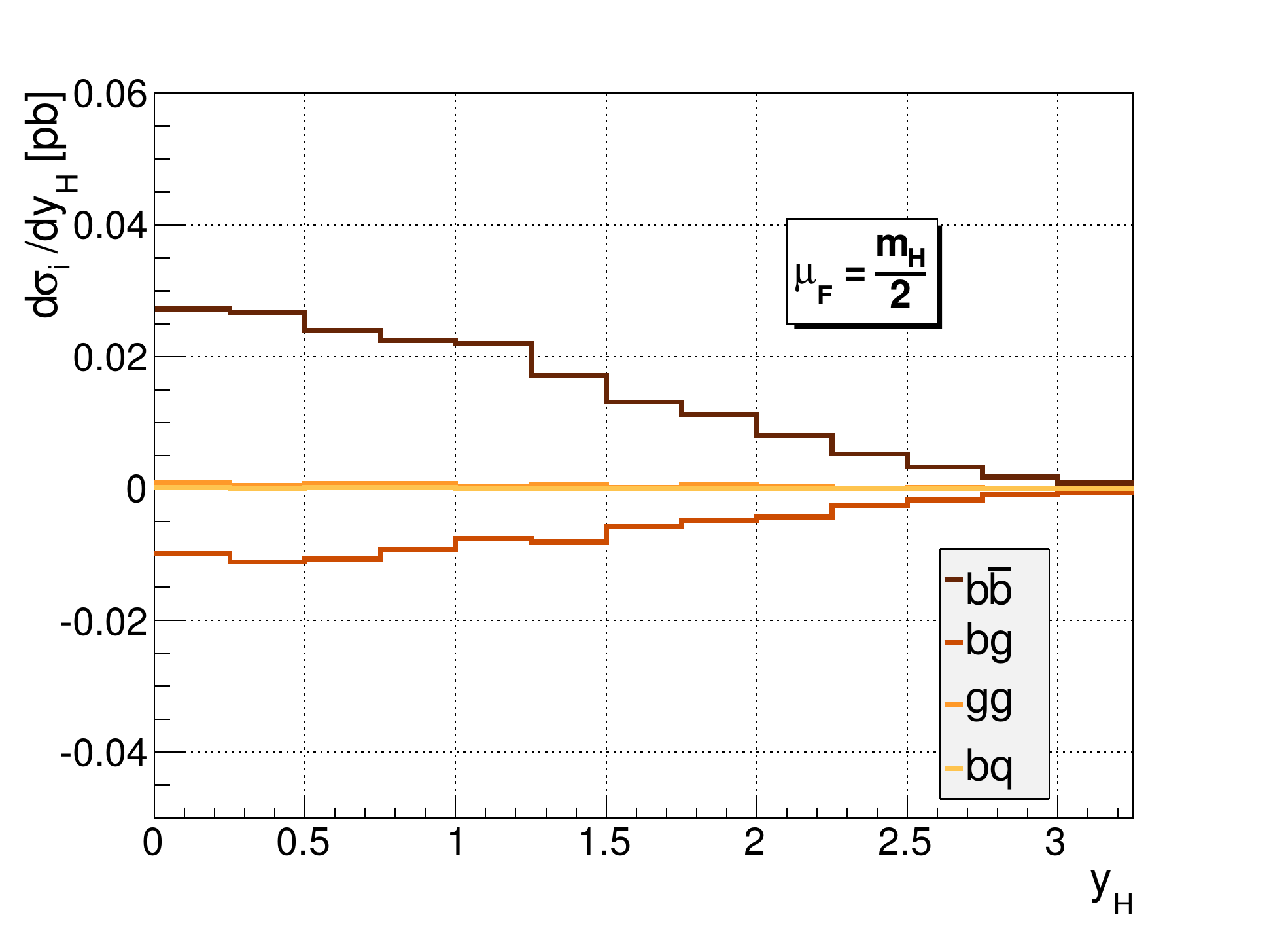}
\end{minipage}
\begin{minipage}[b]{0.49\textwidth}
\includegraphics[width=\textwidth]{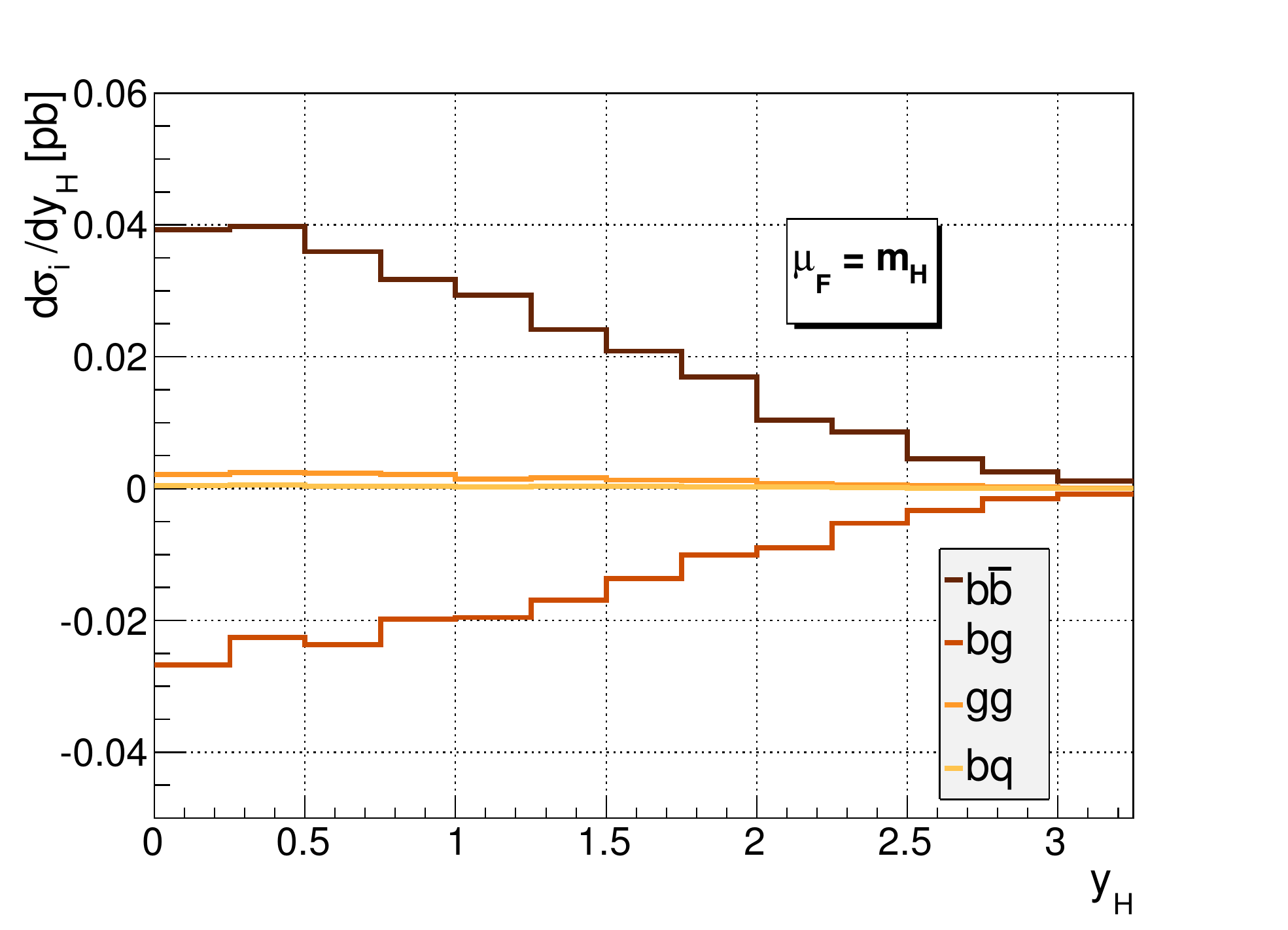}
\end{minipage}
\begin{minipage}[b]{0.49\textwidth}
\includegraphics[width=\textwidth]{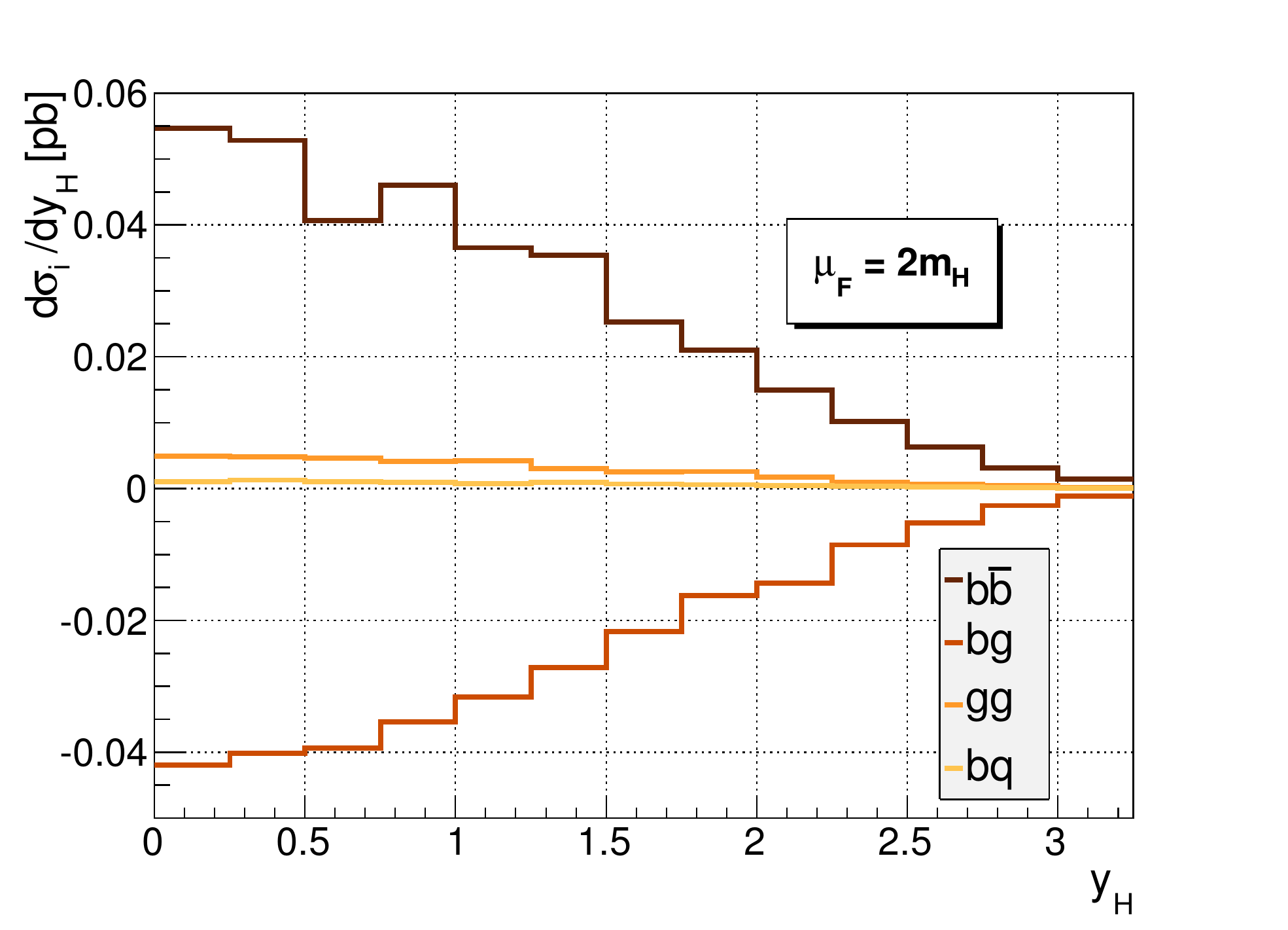}
\end{minipage}
\caption{The distribution of the Higgs absolute rapidity, $|y|$ per initial state channel for $m_H = 125$ GeV at the $8$ TeV LHC, with $\mu_F={m_H\over 16},{m_H\over 8},{m_H\over 4},{m_H\over 2},m_H,2m_H$.}
\label{plot:channelsY}
\end{figure}
\begin{figure}[htbp]
\centering
\begin{minipage}[b]{0.49\textwidth}
\includegraphics[width=\textwidth]{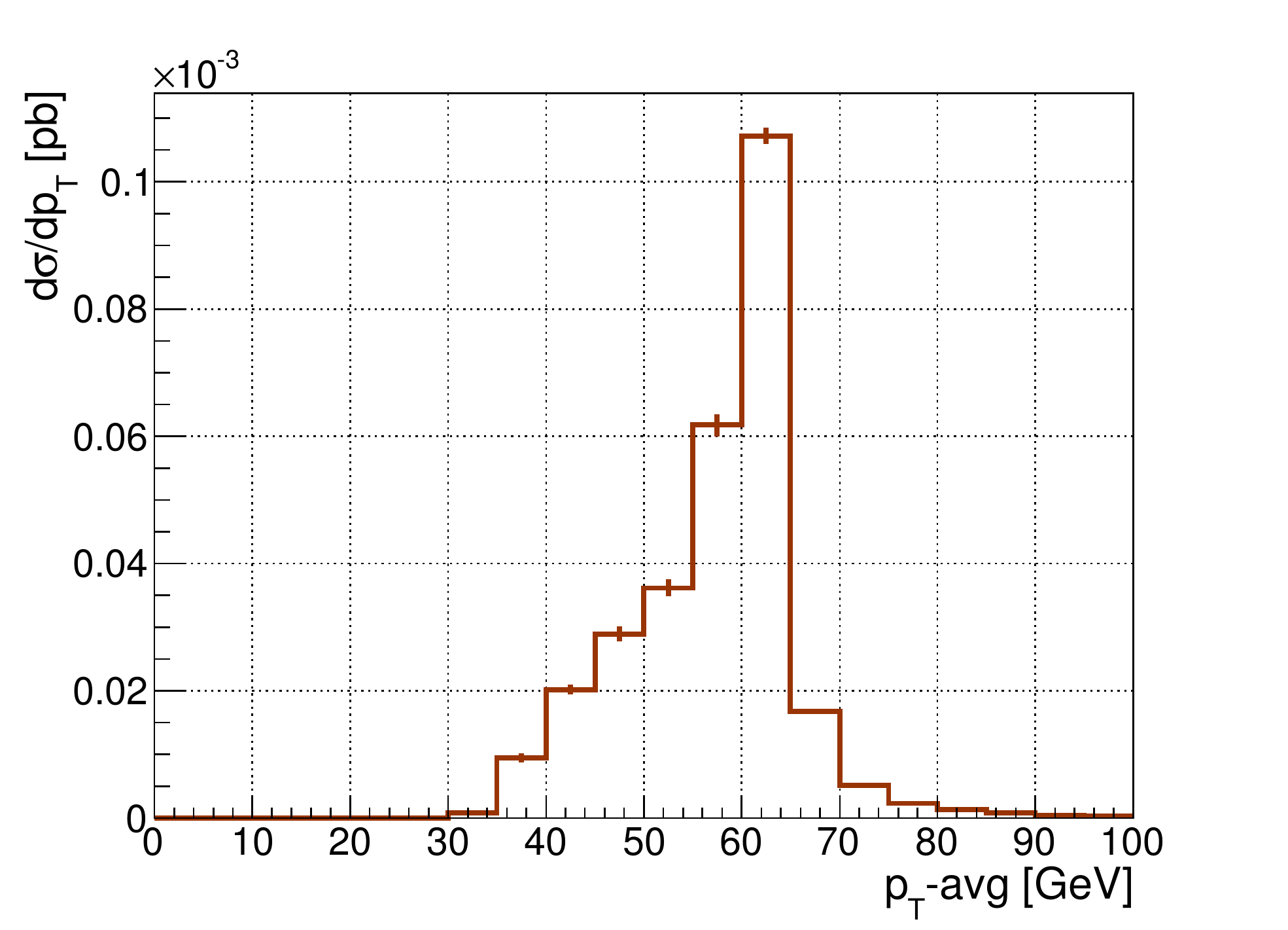}
\end{minipage}
\begin{minipage}[b]{0.49\textwidth}
\includegraphics[width=\textwidth]{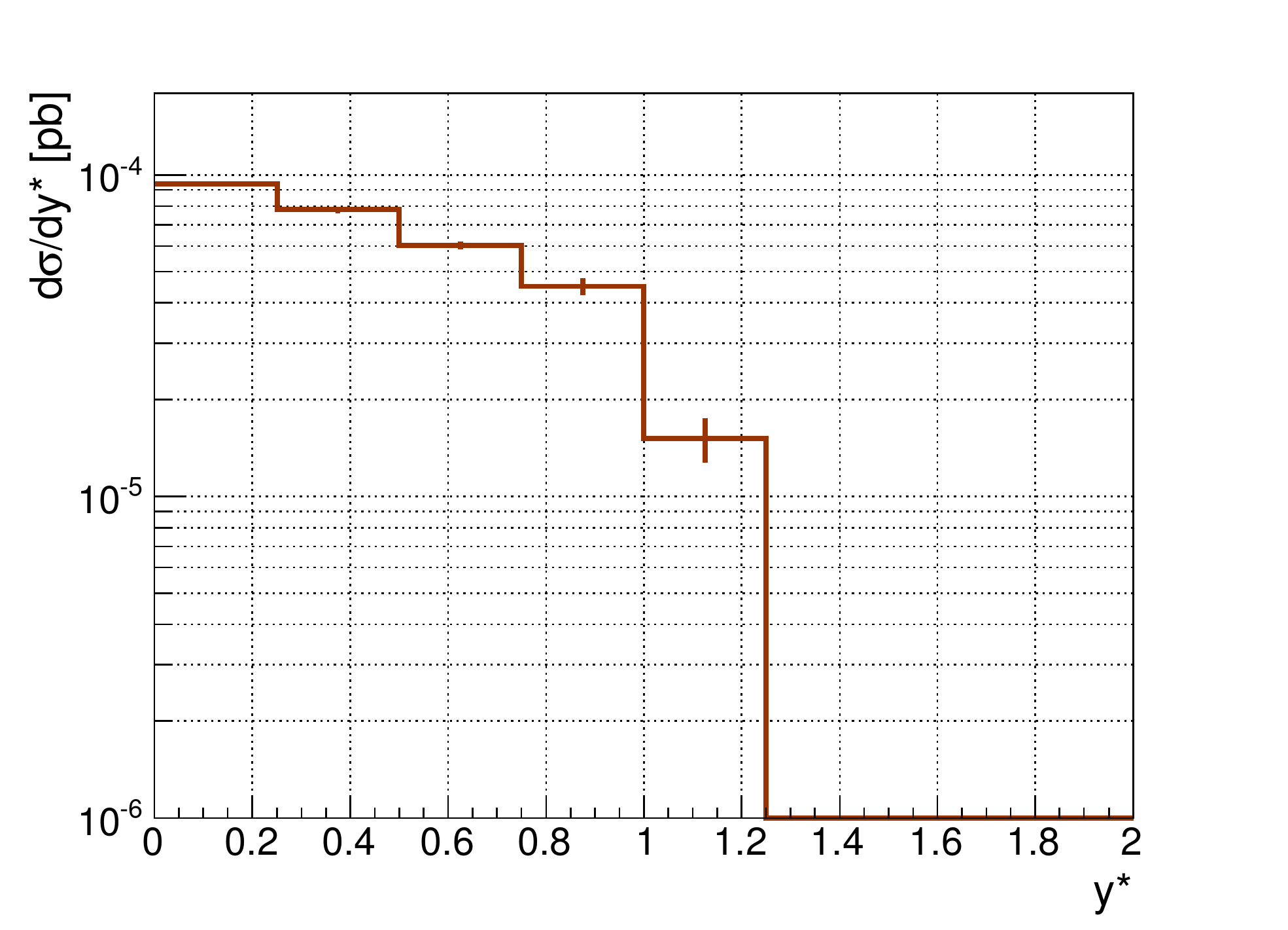}
\end{minipage}
\caption{The average $p_T$ of the two photons
and the $Y^*$ distribution for $b\bar{b}\to H+X\to\gamma\gamma+X$ for $m_H = 125$ GeV at the $8$ TeV LHC, in the presence of cuts described in the text.}
\label{plots:photons}
\end{figure}
A wealth of information can be derived from examining the contribution of the different initial state channels to differential distributions. The six initial state channels that contribute to our NNLO calculation have singularities in various collinear regions that are canceled against the collinear counter terms from mass factorization. In order to make the cross section per channel finite one has to use collinear counter terms that include $\Gamma_{ij}^{(m)}$ kernels involving only the initial state partons of the channel considered.  Since we calculate the collinear counter terms numerically this modification was relatively easy to achieve. 

Initial state channel contributions to differential distributions have a strong dependence on the factorization scale, as do initial state channel contributions to the inclusive cross section. In fig.~\ref{plot:channels_PT} we see the contributions to the Higgs boson $p_T$ distribution from each channel, for various factorization scales ranging from $m_H / 16$ to $2m_H$.

Within the 5FS, the factorization scale regularizes the collinear singularities which in the 4FS are regularized by the bottom mass. At NNLO, three initial state channels, $b\bar{b}$, $bg$ and $gg$ share common collinear configurations whose leading logarithms cancel each other in different bins of the Higgs $p_T$ distribution. In the zero $p_T$ bin, in particular, squared logarithms from the double collinear limit of the $gg$ channel cancel against the single collinear limit of the $bg$ channel and the born contribution of the $b\bar{b}$ channel. Moreover at NNLO one also sees sub-leading (single) logarithms canceling each other between the single collinear configurations of the $gg$ channel and the regular contributions to the $bg$ channel, a cancellation that appears in non-zero $p_T$ bins as well. The magnitude of those logarithmic cancellations is regulated by the value of the factorization scale. The factorization scale dependence is an artifact of the truncation of the perturbative series, so one would naively choose the scale in a way that minimizes the cross-channel logarithmic cancellations. However, choosing the scale too small reduces the regime where the logarithms are re-summed in the PDFs, destabilizing the perturbative expansion. Ideally one should choose the scale in the region where the collinear approximation implicit in the 5FS is still reasonable, which is at $m_H/4$ or lower. 
Corroborative evidence for such a choice comes from the behavior of
the average transverse momentum and of the average rapidity of the
Higgs boson as a function of the factorization scale choice, shown in
fig.~\ref{plot:av_pt_rap}. 

These features are also seen in the rapidity distribution of the Higgs boson per initial state channel, shown in fig.~\ref{plot:channelsY} for various values of $\mu_F$. There it is clearly seen that a scale like $\mu_F=m_H/4$ eliminates the cross-channel cancelations but a lower scale $\mu_F=m_H/16$ leads to a reduced, $bg$-dominated prediction.

We turn now to more exclusive observables.
In  large $\tan{\beta}$ models where the Higgs boson production gets significant contribution from the bottom quark annihilation process, one would like to examine differential distributions involving decay products of the Higgs boson, with cuts necessary in the experimental analyses. We focus here, for demonstration purposes, on the case where the Higgs boson decays to two photons. In such an analysis the minimal cuts used by CMS and ATLAS include:
\begin{itemize}
\item A cut on the $p_T$ of the leading photon: $p_{T;1}>40$GeV.
\item A cut on the $p_T$ of the trailing photon: $p_{T;2}>25$GeV.
\item A cut on the rapidity of both photons: $|y_{1,2}|<2.4$.
\item An isolation cut on photons: no jet is allowed in a cone of radius $0.4$ around any of the two photons if it is $p_T>15$GeV.
\end{itemize}
We treat the Higgs boson in the zero width approximation in this article. We defer a more realistic treatment of the Higgs propagator to future work. 

Within this setup we show in fig.~\ref{plots:photons}  the distribution of the average transverse momentum of the two photons and the distribution of the absolute of the difference in pseudo-rapidity between the two photons, $Y^*=\frac{1}{2}|y_1-y_2|$.


\section{Conclusions}
\label{sec:conclusions}

We have presented the fully differential NNLO calculation of $b\bar{b}\to H$, a process  of prime phenomenological importance for the LHC in all models with enhanced bottom Yukawa couplings. This is the first independent cross-check of the inclusive NNLO calculation performed in~\cite{Harlander:2003ai}. We have presented a variety of differential distributions for Higgs production that can only be obtained with a fully differential calculation and are useful for assessing the quality of the perturbative expansion and the level under which several features are under control at a fully differential level. We have also presented predictions for fully exclusive observables for the $b\bar{b}\to H \to \gamma\gamma$ process in the presence of tight cuts on the final state photons including isolation cuts, demonstrating that our calculation can fully simulate any experimental setup at the partonic level.

This is the second application of our approach to treat real emission singular amplitudes at NNLO~\cite{Anastasiou:2010pw}. It is  the first application for the more complicated case of a hadron collider process. We find the approach particularly beneficial, both in terms of automatization and in terms of performance of the resulting numerical code. We find that the improvement in performance compared to the sector decomposition approach 
is significant. We intend to release the computer code in the near future and we defer for then any detailed comments on performance issues. 

A study of significantly wider scope, including the production via gluon fusion in models with enhanced bottom Yukawa couplings, as well as the decay of Higgs to bottom quarks or tau leptons would vastly benefit the experimental searches. We defer such a study for a future publication.

\section{Acknowledgements}
We  thank Babis Anastasiou for some great ideas and suggestions as well as for independently computing  
the analytic expressions for the threshold limits of the double real and real-virtual partonic cross sections.
This research is supported by the ERC Starting Grant for the project ``IterQCD'' and the Swiss National Foundation under contract SNF 200020-126632.  



\appendix

\section{Threshold behaviour}
\label{Treshholdexplicit}
\subsection{Double real}
The $z$ dependence of the double real soft contribution factorizes completely
\[
	\sigma^{RR}_S =  4\, C_F \, \mathcal{B} \, \mathrm{e} ^ {- 2 l_H \epsilon}\,(1-z)^{-1-4 \epsilon}\, \Delta_{RR}^{(4)} \, ,
\]
where $l_H$ is defined in eq.(\ref{lH}), $\mathcal{B}$ in eq.(\ref{equ:B}) and
\begin{align*}
	\Delta_{RR}^{(4)} & = \left( \frac 1 4 \epsilon^{-3} - \frac 7 4 \zeta_2 \epsilon^{-1} - \frac{31} {6} \zeta_3  - \frac 9 {16} \zeta_4 \, \epsilon \right) C_A^{-1} \\
	& \qquad + \left( \frac 1 {24} \epsilon^{-2} + \frac {5}{72} \epsilon^{-1} + \frac {7}{54} - \frac {7}{24} \zeta_2 + \left( \frac {41}{162} - \frac{35}{72} \zeta_2 - \frac{31}{36} \zeta_3 \right) \epsilon \right) n_f \\
	& \qquad - \left(\frac 3 8 \epsilon^{-3} +\frac{11}{48} \epsilon^{-2} + \left(\frac {67}{144} - \frac{11}{4} \zeta_2 \right) \epsilon^{-1} + \frac {101}{108}-\frac{77}{48} \zeta_2 -\frac{67}{8} \zeta_3 \right. \\
	& \qquad \qquad \left. + \left( \frac{607}{324} - \frac{469}{144} \zeta_2 - \frac{341}{72} \zeta_3 - \frac{19}{32} \zeta_4 \right) \epsilon \right) C_A  + \mathcal{O}(\epsilon^2) \,.
\end{align*}

\subsection{Real-virtual}
We decompose the real-virtual soft contribution as
\[
	\sigma^{RV}_S =4\, C_F \, \mathcal{B} \, \mathrm{e} ^ {- 2 l_H \epsilon}\, \sum_{n} (1-z)^{-1-n \epsilon} \Delta_{RV}^{(n)} \, ,
\]
where only $\Delta_{RV}^{(2)} $ and $\Delta_{RV}^{(4)} $ are non vanishing and given by
\begin{align}
	\Delta_{RV}^{(2)} &= 2\, C_F \left( \frac{1}{4} \, \epsilon^{-3} + \left( \frac{1}{4}- \frac{5}{4} \zeta_2 \right) \epsilon^{-1} + \frac{1}{2}-\frac{7}{6} \zeta _3 + \left( 1-\frac {5}{4} \zeta_2 +\frac {67}{16} \zeta_4 \right) \epsilon \right) + \mathcal{O}(\epsilon^2) \, , \\
	\Delta_{RV}^{(4)} &= C_A \left( \frac{1}{8}\, \epsilon^{-3}-\frac{7}{8} \zeta_2 \, \epsilon^{-1} - \frac{7}{3} \zeta_3 -\frac {21}{32} \zeta_4\, \epsilon \right) + \mathcal{O}(\epsilon^2)\, .
\end{align}

\section{Scale separation}
\label{scaleseparation}
The renormalization and factorization scales, $\mu_R$ and $\mu_F$, can be conveniently separated by 
first setting $\mu=\mu_F$ and then applying the following relations 
\begin{eqnarray}
\frac{\alpha_s(\mu_F)}{\pi}  & = & \frac{\alpha_s(\mu_R)}{\pi}+  \left(\frac{\alpha_s(\mu_R)}{\pi}\right)^2 \beta_0 \log\left(\frac{\mu_R^2}{\mu_F^2}\right) \nonumber \\ 
& &  + \left(\frac{\alpha_s(\mu_R)}{\pi}\right)^3 \left[\beta_1 \log\left(\frac{\mu_R^2}{\mu_F^2}\right)+\beta_0^2 \log^2\left(\frac{\mu_R^2}{\mu_F^2}\right)\right]+\mathcal{O}\left(\alpha_s^4\right) \, ,\nonumber \\
y_b(\mu_F)& = & y_b(\mu_R) \Bigg\lbrace  1+  \frac{\alpha_s(\mu_R)}{\pi} \gamma_0 \log\left(\frac{\mu_R^2}{\mu_F^2}\right)\nonumber  \\
& &+\left(\frac{\alpha_s(\mu_R)}{\pi}\right)^2 \left[\gamma_1 \log\left(\frac{\mu_R^2}{\mu_F^2}\right)+\frac{1}{2}(\gamma_0\beta_0+\gamma_0^2)\log^2\left(\frac{\mu_R^2}{\mu_F^2}\right)\right]
+\mathcal{O}\left(\alpha_s^3\right) \Bigg\rbrace
\end{eqnarray}
where 
\begin{eqnarray}
\gamma_0 &=& 1, \qquad \: \quad \qquad \gamma_1 = \frac{101}{24}-\frac{5n_f}{36}, \nonumber \\ 
\beta_0 &=&  \frac{11}{4}-\frac{n_f}{6}, \qquad \beta_1 =  \frac{51}{8}-\frac{19n_f}{24}. 
\end{eqnarray}



\providecommand{\href}[2]{#2}
\begingroup
\raggedright
\endgroup

\end{document}